\documentclass[12pt,draftcls,onecolumn]{IEEEtran}
\usepackage{amsmath,amsfonts}
\usepackage{multirow}
\usepackage{algorithm}
\usepackage{xcolor}
\usepackage{algcompatible}
\usepackage{array}
\usepackage[caption=false,font=normalsize,labelfont=sf,textfont=sf]{subfig}
\usepackage{textcomp}
\usepackage{stfloats}
\usepackage{url}
\usepackage{verbatim}
\usepackage{graphicx}
\usepackage{cite}
\usepackage{breqn}
\usepackage{hyperref}

\usepackage{environ}         
\usepackage{etoolbox}        
\usepackage{graphicx}        

\begin{document}
\title{Dual-Polarized IRS-Assisted MIMO Network}
\author{Muteen~Munawar
        and Kyungchun~Lee,~\IEEEmembership{Senior Member, IEEE}
    \thanks{This work was supported in part by the Basic Science Research Program through the National Research Foundation of Korea (NRF) funded by the Ministry of Education under Grant NRF-2019R1A6A1A03032119 and in part by the NRF Grant funded by the Korean Government (MSIT) under Grant NRF-2022R1A2C1006566. (Corresponding author: Kyungchun Lee.) }
    \thanks{M.~Munawar and K.~Lee are with the Department of Electrical and Information Engineering and the Research Center for Electrical and Information Technology, Seoul National University of Science and Technology, Seoul 01811, Republic of Korea (e-mail: muteen@seoultech.ac.kr, kclee@seoultech.ac.kr).}
}
\maketitle
\vspace{-1cm}

\begin{abstract}
This study considers a dual-polarized intelligent reflecting surface (DP-IRS)-assisted multiple-input multiple-output (MIMO) single-user wireless communication system. The transmitter and receiver are equipped with DP antennas, and each antenna features a separate phase shifter for each polarization. We attempt to maximize the system's spectral efficiency (SE) by optimizing the operations of the reflecting elements at the DP-IRS, precoder/combiner at the transmitter/receiver, and vertical/horizontal phase shifters at the DP antennas. To address this problem, we propose a three-step alternating optimization (AO) algorithm based on the semi-definite relaxation method. Next, we consider asymptotically low/high signal-to-noise ratio (SNR) regimes and propose low-complexity algorithms. In particular, for the low-SNR regime, we derive computationally low-cost closed-form solutions. According to the obtained numerical results, the proposed algorithm outperforms the various benchmark schemes. Specifically, our main algorithm exhibits a 65.6 \% increase in the SE performance compared to random operations. In addition, we compare the SE performance of DP-IRS with that of simple IRS (S-IRS). For \(N = 50\), DP-IRS achieves 24.8 \%, 28.2 \%, and 30.3 \% improvements in SE for \({4} \times {4}\), \({8} \times {8}\), and \({16} \times {16}\) MIMO, respectively, compared to S-IRS.
\end{abstract}

\section{Introduction}
\label{Sect:Intro}
An intelligent reflecting surface (IRS) is a metasurface \cite{dai2021wireless} with several reflecting elements of the passive nature \cite{ozdogan2019intelligent}. By applying a bias voltage, we can alter the material characteristics of these elements \cite{abeywickrama2020intelligent}, resulting in a change in the amplitudes and phases of the impinging signals. Owing to its  real-time configuration \cite{cui2014coding}, IRS is considered a potential candidate for next-generation wireless communication systems \cite{gong2020toward}. In the literature, the IRS has been used to improve the signal-to-noise ratio (SNR) \cite{abeywickrama2020intelligent, wu2019intelligent, wu2019beamforming}, increase security \cite{chen2019intelligent,cui2019secure,dong2020secure,yang2020deep,feng2020physical}, reduce interference \cite{wu2019intelligent,jia2020analysis,jiang2021achievable}, transfer power \cite{pan2020intelligent,el2020performance}, and reflect modulation \cite{lin2020reconfigurable}. In addition, fast channel estimation methods and low-complexity beamforming schemes for IRS have been studied in the literature \cite{PP2published,zheng2022survey,wang2020channel,jiang2021learning}.

The introduction of dual polarization (DP) in metasurfaces can further increase their capabilities with polarization diversity (PD) \cite{vaughan1990polarization,lee1972polarization}, polarization multiplexing (PM) \cite{deng2020malus}, and polarization modulation \cite{krongold2011comparison}. Although modern communication systems operate with dual-polarized waves, to the best of our knowledge, only a few studies exist on the application of DP-IRS. For instance, in \cite{de2021irs}, the authors adopted DP-IRS for PM, to reduce the interference problem in the system. Furthermore, in \cite{chen2021design}, the authors exploited the orthogonal property of DP-IRS to transmit two data streams by modulating the data directly on the reflecting elements. Based on the inherent properties of dual polarization, there is a need to further explore the potential benefits of DP-IRS. 

Here, we consider a DP-IRS-assisted narrow-band multiple-input multiple-out (MIMO) wireless communication network, where the transmitter, i.e., access point (AP), and receiver, i.e., user, are equipped with multiple DP antennas \cite{shu2018dual}. Each DP antenna is equipped with a single RF chain and a phase shifter for each polarization \cite{arai2013dual}. We attempt to maximize the spectral efficiency of the DP-IRS-assisted MIMO network by optimizing the operation of the DP-IRS. Only a few studies exist on capacity maximization for IRS, which are limited to simple IRS (S-IRS)-assisted MIMO networks \cite{bjornson2019intelligent}. For example, in \cite{zhang2020capacity}, the authors optimized the reflecting elements of the S-IRS, where the transmitter and receiver were equipped with single polarized antennas. Similarly, in \cite{9777870}, the authors considered an S-IRS-assisted MIMO network and exploited millimeter-wave channel characteristics to devise low-complexity beamforming methods and maximize the achievable rate. However, to simplify the problem in \cite{9777870}, the authors assumed a weak AP-user channel and only considered the AP-IRS-user link in the optimization problem. In our proposed technique, we consider DP-IRS and assume both direct and reflected links in the problem. Our main contributions are summarized as follows.

\begin{itemize}
\item We study the DP-IRS-assisted MIMO wireless network and formulate a capacity maximization problem, which is non-convex. To solve this problem, we decompose the problem into several sub-problems and propose a semi-definite relaxation (SDR)-based alternating optimization (AO) algorithm. The algorithm works by optimizing the operation of the reflecting elements at the DP-IRS, precoder/combiner matrices at the AP/user, and vertical/horizontal phase shifters at DP antennas alternatively, until the spectral efficiency converges. The algorithm and problem formulations are designed such that the objective value stays non-decreasing and is guaranteed to converge.
\item Next, we consider two special cases: 1) low-SNR regime and/or line of sight (LoS) scenario and 2) high-SNR regime and rich scattered environment. Case 1 occurs when the user is in the low-SNR regime and/or the dominant channel between the AP and the user is LoS. The LoS channel occurs when the IRS is mounted in the AP’s line of sight. Similarly, high-SNR and rich-scattered environments favor Case 2. We propose two low-complexity AO algorithms to maximize the spectral efficiency for these two special cases. Specifically, for Case 1, we derive closed-form solutions.
\item Finally, we provide an extensive numerical analysis of the system to demonstrate the proposed algorithms' convergence behaviors and SE performances.

\end{itemize}

The remainder of this paper is organized as follows. Section II discusses the system model and formulates the capacity-maximization problem. The proposed algorithms are presented in Sections III and IV. In Section V, we numerically evaluate the performance of the proposed algorithms and compare them with benchmark schemes. Finally, Section VI concludes the paper.

\textit{Notations:} Scalars are denoted by italic letters, whereas vectors and matrices are denoted by bold-face lower- and upper-case letters, respectively. The subscripts/superscripts \(v\) and \(h\) denote vertical and horizontal polarizations, respectively. For a complex-valued vector \(\bf{v}\) of length \(K\), \({\left(  \bf{v}  \right)^T}\) denotes the transpose, \({{\bf{v}}_k}\) denotes the \(k\)th element of \(\bf{v}\), \({{\bf{v}}^*}\) denotes the complex conjugate of each element, \({{\bf{v}}^H}\) denotes the conjugate transpose, \(\left\| {\bf{v}} \right\|\) denotes the Euclidean norm, \({\rm{diag}}\left( {\bf{v}} \right)\) denotes a diagonal matrix with each diagonal element being the corresponding element in \(\bf{v}\), \(\arg ({\bf{v}})\) denotes a vector with each element being the phase of the corresponding element in \(\bf{v}\), \({\rm{logsum}}\left( {\bf{v}} \right)\) means \(\sum\limits_{k = 1}^K {{{\log }}\left( {{\bf{v}}}_k \right)} \), \({\rm{sum}}\left( {\bf{v}} \right)\) means \(\sum\limits_{k = 1}^K {\left( {{\bf{v}}}_k \right)} \), and \({\rm{GM}}\left( {\bf{v}} \right)\) represents the geometric mean. For a complex-valued matrix \({\bf{M}}\), \({\rm{rank}}\left( {\bf{M}} \right)\) denotes the rank, \({{\bf{M}}_{\left( {n,m} \right)}}\) denotes the \(n\)th row and \(m\)th column entry, \({{{\bf{M}}^\dag }}\) denotes the pseudoinverse, \(\left\| {\bf{M}} \right\|\) denotes the Frobenius norm, \({\bf{M}}\underline  \succ  0\) denotes positive semi-definite, \({\rm{trace}}\left( {\bf{M}} \right)\) denotes trace, \({\bf{M}} \circ {\bf{M}}\) denotes the Hadamard product, and \(\det \left( {\bf{M}} \right)\) denotes the determinant. For a complex-valued square matrix \({\bf{A}}\), \({{\bf{I}}_{\bf{A}}}\) denotes an identity matrix of size \({\bf{A}}\). The notation \({\left(  \cdot  \right)^*}\) denotes the complex conjugate of a complex number.

\section{System Model}

\label{Sect:SystemModel}
\begin{figure}[!h]
\centering
\includegraphics[width=0.8\textwidth]{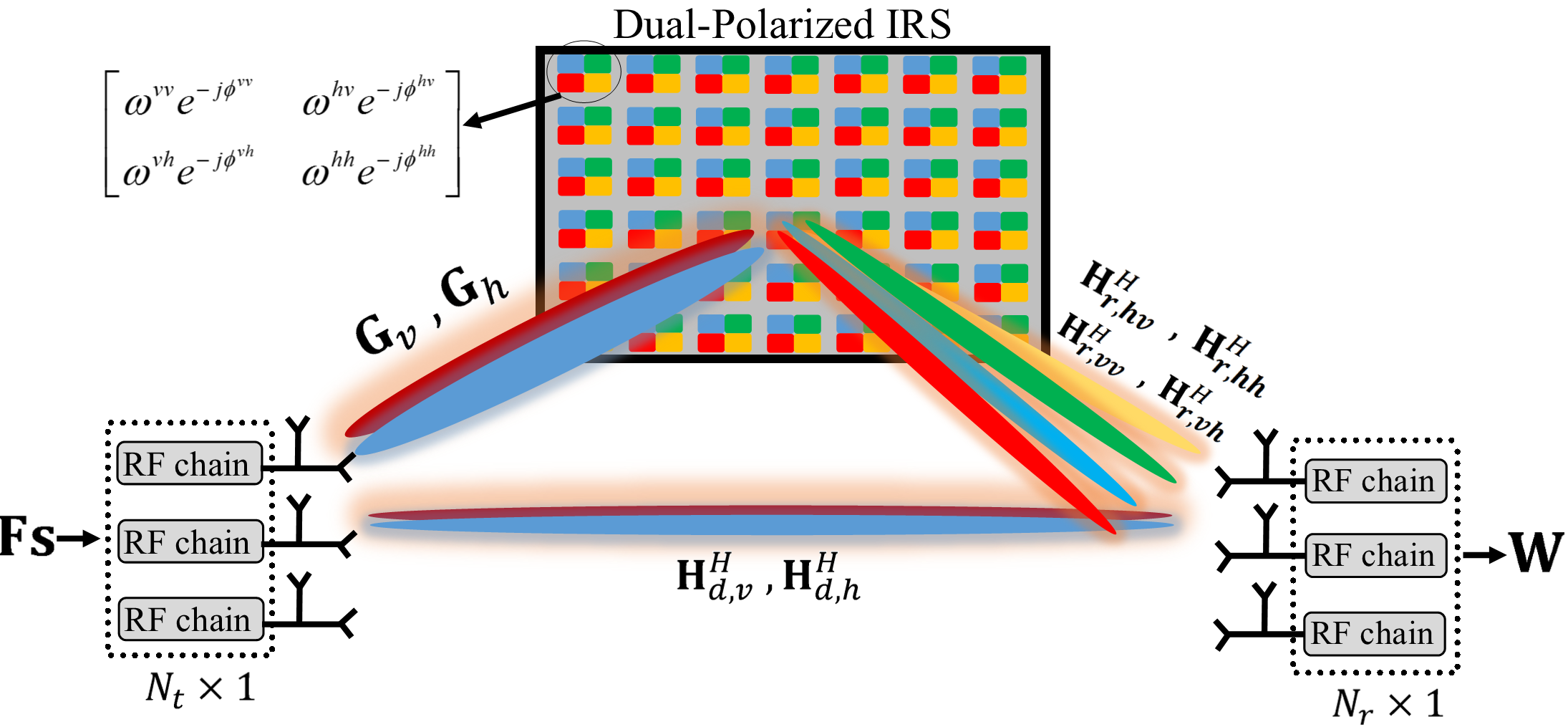}
\caption{\textcolor{black}{System model.}}
\label{fig:SystemModel}
\end{figure}
We consider a DP-IRS-aided MIMO wireless communication network, as illustrated in Fig. \ref{fig:SystemModel}. The AP and user are equipped with \(N_t\) and \(N_r\) DP antennas, respectively, and each DP antenna is connected to a single RF chain but with separate phase shifters for each polarization \cite{arai2013dual,shu2018dual}. DP-IRS comprises \(N\) passive reflecting elements, and the operation of each element can be defined by the following transformation matrix \cite{de2021irs}:
\begin{equation} \label{eq1}
{\bf{\Psi }} = \left[ {\begin{array}{*{20}{c}}
{{\omega ^{vv}}{e^{ - j{\phi ^{vv}}}}}&{{\omega ^{hv}}{e^{ - j{\phi ^{hv}}}}}\\
{{\omega ^{vh}}{e^{ - j{\phi ^{vh}}}}}&{{\omega ^{hh}}{e^{ - j{\phi ^{hh}}}}}
\end{array}} \right],
\end{equation}
\textcolor{black}{where \({\omega ^{pq}} \in \left[ {0,1} \right]\) and \({\phi ^{pq}} \in \left[ {0,2\pi } \right)\) represent the amplitude and phase changes of the reflected signal from polarization \(p\) to \(q\), respectively, and \(p,q \in \{ v,h\} \).} 
\textcolor{black}{The transformation matrix \eqref{eq1} elucidates that the DP-IRS, in addition to possessing the ability for phase and amplitude control, which is also a feature of S-IRS, can effectively execute polarization beam splitting, independent control of incident polarization, and polarization conversion. The attributes and characteristics of these DP-IRSs have been thoroughly investigated in the domains of antennas and electromagnetic theory, as evidenced by the works of \cite{saha21,saha22,saha23}. Moreover, their potential has been exploited in the context of non-orthogonal multiple access networks \cite{de2021irs}}.

\textcolor{black}{The received signal at the user is a linear superposition of two signals originating from the AP and DP-IRS, respectively. Considering only a single antenna and a solitary reflecting element at the AP/user and DP-IRS, respectively, the aggregate noise-free signal received by the user is expressed as}
\begin{equation} \label{eq2} \color{black} {
\begin{array}{l}
\left[ {\begin{array}{*{20}{c}}
{{y^v}}\\
{{y^h}}
\end{array}} \right] = \left( {\begin{array}{*{20}{c}}
{{e^{ - j{\gamma ^v}}}}&0\\
0&{{e^{ - j{\gamma ^h}}}}
\end{array}} \right)\left[ {\left\{ {\left( {\begin{array}{*{20}{c}}
{{{\left( {h_r^{vv}} \right)}^*}}&{{{\left( {h_r^{hv}} \right)}^*}}\\
{{{\left( {h_r^{vh}} \right)}^*}}&{{{\left( {h_r^{hh}} \right)}^*}}
\end{array}} \right)} \right.} \right.\\
\,\,\,\,\,\,\,\,\,\,\,\, \circ \underbrace {\left( {\begin{array}{*{20}{c}}
{{\omega ^{vv}}{e^{ - j{\phi ^{vv}}}}}&{{\omega ^{hv}}{e^{ - j{\phi ^{hv}}}}}\\
{{\omega ^{vh}}{e^{ - j{\phi ^{vh}}}}}&{{\omega ^{hh}}{e^{ - j{\phi ^{hh}}}}}
\end{array}} \right)}_{\bf{\Psi }} \circ \left( {\begin{array}{*{20}{c}}
{{g^{vv}}}&{{g^{hv}}}\\
{{g^{vh}}}&{{g^{hh}}}
\end{array}} \right)\\
{\left. {{{\left. {\, + \left( {\begin{array}{*{20}{c}}
{{{\left( {h_d^v} \right)}^*}}&0\\
0&{{{\left( {h_d^h} \right)}^*}}
\end{array}} \right)} \right\}}^T}\left( {\begin{array}{*{20}{c}}
{\sqrt {\frac{{{P_t}}}{2}} {e^{ - j{\theta ^v}}}}&0\\
0&{\sqrt {\frac{{{P_t}}}{2}} {e^{ - j{\theta ^h}}}}
\end{array}} \right)} \right]^T}x,
\end{array}}
\end{equation}
where \({{y^v}}\), \({{y^h}}\), \({(h_r^{})^*}\), \({(h_d^{})^*}\), \(g\), and \(x\) denote the received vertical polarized signal, received horizontal polarized signal, IRS-user channel, AP-user channel, AP-IRS channel, and transmitted symbol, respectively.
The matrices \({\rm{diag}}\left( {\frac{P_t}{2}{e^{ - j{\theta ^v}}},\frac{P_t}{2}{e^{ - j{\theta ^h}}}} \right)\) and \({\rm{diag}}\left( {{e^{ - j{\gamma ^v}}},{e^{ - j{\gamma ^h}}}} \right)\) in \eqref{eq2} represent the precoder at the AP and combiner at the user, respectively. \textcolor{black}{In \eqref{eq2}, because of the single RF chain, the total transmission power \(P_t\) is divided equally between the two radiated polarizations. Subsequently, the polarization signals are combined at the receive DP antenna as}
\begin{equation} \label{eq2a} \tag{2a} \color{black}
\begin{array}{l}
y = {y^v} + {y^h}\\
\,\,\,\, = \left( {{e^{ - j{\gamma ^v}}}\left\{ {{{\left( {h_r^{vv}} \right)}^*}{\omega ^{vv}}{e^{ - j{\phi ^{vv}}}}{g^{vv}} + {{\left( {h_r^{hv}} \right)}^*}{\omega ^{hv}}{e^{ - j{\phi ^{hv}}}}{g^{hv}} + {{\left( {h_d^v} \right)}^*}} \right\}\frac{{{P_t}}}{2}{e^{ - j{\theta ^v}}}x} \right)\\
\,\,\,\,\,\,\, + \left( {{e^{ - j{\gamma ^h}}}\left\{ {{{\left( {h_r^{vh}} \right)}^*}{\omega ^{vh}}{e^{ - j{\phi ^{vh}}}}{g^{vh}} + {{\left( {h_r^{hh}} \right)}^*}{\omega ^{hh}}{e^{ - j{\phi ^{hh}}}}{g^{hh}} + {{\left( {h_d^h} \right)}^*}} \right\}\frac{{{P_t}}}{2}{e^{ - j{\theta ^h}}}x} \right),
\end{array}
\end{equation}
\textcolor{black}{where \({\left( {h_r^{pp}} \right)^*}{\omega ^{pp}}{e^{ - j{\phi ^{pp}}}}{g^{pp}}\), \(p \in \left\{ {v,h} \right\}\), represents the total channel gain of the signal, which traverses through the polarization \(p\) of the AP-IRS link and is subsequently reflected towards the polarization \(p\), \(p \in \left\{ {v,h} \right\}\), of the IRS-User link. This reflection occurs without any polarization conversion and is then focused on the corresponding \(p\), \(p \in \left\{ {v,h} \right\}\), component of the receive DP antenna. In contrast, in \eqref{eq2}, \({\left( {h_r^{pq}} \right)^*}{\omega ^{pq}}{e^{ - j{\phi ^{pq}}}}{g^{pq}}\), \(p,q \in \left\{ {v,h} \right\}\), \(p \ne q\), represents the total channel gain in the reflected path involving the conversion of the signal, which traverses through the polarization \(p\) to the polarization \(q\). This reflected signal is directed towards the polarization \(q\) of the IRS-User link.}

\textcolor{black}{It is noteworthy that \eqref{eq2a} pertains to a single-input single-output system, wherein a single transmit/receive DP antenna is accompanied by a single RF chain. In such a configuration, the deployment of DP-IRS and DP antennas only achieves the PD gains, while providing no PM gains. Nonetheless, upon extending this model to encompass MIMO scenarios, where an array of DP antennas and multiple data streams are available, the beam splitting and polarization conversion capabilities of DP-IRS supplement the PD gains. This facilitates more effective control of the split orthogonal beams in the IRS-User links and the beams emanating from the AP-User links; consequently resulting in the interlayer multiplexing gains.} 

The signal model given in \eqref{eq2a} can be extended to \(N_t\) transmit antennas, \(N_r\) receive antennas, and \(N\) reflecting elements as
\begin{equation} \label{eq3} \color{black}
\begin{array}{l}
{\bf{y}} = {\bf{W}}\Bigr[ {{{\bf{E}}_v}\left\{ {{\bf{H}}_{r,vv}^H{{\bf{\Theta }}_{vv}}{{\bf{G}}_{vv}} + {\bf{H}}_{r,hv}^H{{\bf{\Theta }}_{hv}}{{\bf{G}}_{hv}} + {\bf{H}}_{d,v}^H} \right\}{{\bf{U}}_v}}\\
{\,\,\,\,\,\,\,\,\,\, + {{\bf{E}}_h}\left\{ {{\bf{H}}_{r,hh}^H{{\bf{\Theta }}_{hh}}{{\bf{G}}_{hh}} + {\bf{H}}_{r,vh}^H{{\bf{\Theta }}_{vh}}{{\bf{G}}_{vh}} + {\bf{H}}_{d,h}^H} \right\}{{\bf{U}}_h}} \Bigr]{\bf{Fs}} + {\bf{Wz}},
\end{array}
\end{equation}
where \({\bf{z}}\) and \({\bf{s}}\) represent a circularly symmetric complex Gaussian noise vector at the user and the transmitted symbol vector at the AP, respectively. In addition, \({\mathbf{F}} \in {\mathbb{C}^{{N_t} \times {N_s}}}\) and \({\mathbf{W}} \in {\mathbb{C}^{{N_r} \times {N_s}}}\) denote the precoding and combining matrices at the AP and user, respectively. \textcolor{black}{The matrices \({{\mathbf{E}}_v} \in {\mathbb{C}^{{N_r} \times {N_r}}}\), \({{\mathbf{E}}_h} \in {\mathbb{C}^{{N_r} \times {N_r}}}\), \({{\mathbf{U}}_v} \in {\mathbb{C}^{{N_t} \times {N_t}}}\), and \({{\mathbf{U}}_h} \in {\mathbb{C}^{{N_t} \times {N_t}}}\) contain the receive vertical, receive horizontal, transmit vertical, transmit horizontal phase shifters in the diagonal, respectively, i.e., \({{\bf{E}}_p} = {\rm{diag}}({e^{ - j{\gamma _{p,1}}}},{e^{ - j{\gamma _{p,2}}}},{e^{ - j{\gamma _{p,3}}}}, \cdots ,{e^{ - j{\gamma _{p,{N_r}}}}})\) and \({{\bf{U}}_p} = {\rm{diag}}\left( {{e^{ - j{\theta _{p,1}}}},{e^{ - j{\theta _{p,2}}}},{e^{ - j{\theta _{p,3}}}}, \cdots ,{e^{ - j{\theta _{p,{N_t}}}}}} \right)\), where \(p \in \left\{ {v,h} \right\}\).}
The baseband equivalent channels of the AP-IRS, IRS-User, and AP-User links are denoted as \({\mathbf{G}} \in {\mathbb{C}^{N \times {N_t}}}\), \({\mathbf{H}}_r^H \in {\mathbb{C}^{{N_r} \times N}}\), and \({\mathbf{H}}_d^H \in {\mathbb{C}^{{N_r} \times {N_t}}}\), respectively.
 
 \textcolor{black}{Note that, in addition to the allocation of the optimal power to each DP antenna, which is further equally divided between two polarizations, \({\bf{F}}\) is employed as a common beamformer for both the polarizations of each DP antenna. Nevertheless, when combined with the transmit phase shifters, it generates a different effective beamformer \({{\bf{U}}_v}{\bf{F}} \) and \({{\bf{U}}_h}{\bf{F}} \) for each polarization. Similarly, the operation of \({\bf{W}}\) combined with the receive phase shifters of each polarization, \({\bf{W}}{{\bf{E}}_v}\) and \({\bf{W}}{{\bf{E}}_h}\), can be conceptualized.}
 
Owing to the same spatial location of each subpart of a reflecting element, we can assume the same channel between identical polarizations of the AP, DP-IRS, and user \cite{CLERCKX201359}. For example, the signals reflected by the \(vv\) and \(hv\) subparts of a reflecting element are directed towards the vertical parts of the receiver; hence, we can assume \({\mathbf{H}}_{r,vv} = {\text{ }}{\mathbf{H}}_{r,hv} = {\mathbf{H}}_{r,v}\) \cite{CLERCKX201359}. Similarly, we can assume \({\mathbf{H}}_{r,hh} = {\text{ }}{\mathbf{H}}_{r,vh} = {\mathbf{H}}_{r,h}\) for the IRS-User link
and \({{\mathbf{G}}_{vv}} = {{\mathbf{G}}_{vh}} = {{\mathbf{G}}_v}\) and \({{\mathbf{G}}_{hh}} = {{\mathbf{G}}_{hv}} = {{\mathbf{G}}_h}\) for AP-IRS link. 
For the proposed work, we assume that all the elements exhibit complete reflection, which can be attained by setting \({\omega ^{vv}} = {\omega ^{hv}} = {\omega ^{vh}} = {\omega ^{hh}} = 1\) \cite{wu2019intelligent,wu2019beamforming}. The signal model in \eqref{eq3} is simplified as
\begin{equation} \label{eq3a} \tag{3a} \color{black}
\begin{array}{l}
{\bf{y}} = {\bf{W}}\left[ {{{\bf{E}}_v}\left\{ {{\bf{H}}_{r,v}^H\left( {{{\bf{\Theta }}_{vv}}{{\bf{G}}_v} + {{\bf{\Theta }}_{hv}}{{\bf{G}}_h}} \right) + {\bf{H}}_{d,v}^H} \right\}{{\bf{U}}_v}} \right.\\
\left. {\,\,\,\,\,\,\,\, + {{\bf{E}}_h}\left\{ {{\bf{H}}_{r,h}^H\left( {{{\bf{\Theta }}_{hh}}{{\bf{G}}_h} + {{\bf{\Theta }}_{vh}}{{\bf{G}}_v}} \right) + {\bf{H}}_{d,h}^H} \right\}{{\bf{U}}_h}} \right]{\bf{Fs}} + {\bf{Wz}}.
\end{array}
\end{equation}
\textcolor{black}{To streamline our analysis and avoid complex mathematical formulations, this study does not consider the optimization of the transmit DP antennas phase shifters \(\left( {{{\bf{U}}_v},{{\bf{U}}_h}} \right)\). Instead, we focus on the optimization of the DP-IRS's beam splitting and polarization conversion capabilities. To this end, we assume \({{\bf{\Theta}} _{vv}} = {{\bf{\Theta}} _{hv}}\) and \({{\bf{\Theta}} _{hh}} = {{\bf{\Theta}} _{vh}}\), which simplifies \eqref{eq3a} as}
\begin{equation} \label{eq4} \color{black}
\begin{array}{l}
{\bf{y}} = {\bf{W}}\Bigr[ {{{\bf{E}}_v}\left\{ {{\bf{H}}_{r,v}^H{{\bf{\Theta }}_{hv}}{{\bf{G}}_v} + {\bf{H}}_{r,v}^H{{\bf{\Theta }}_{hv}}{{\bf{G}}_h} + {\bf{H}}_{d,v}^H} \right\}} \\
 {\,\,\,\,\,\,\,\,\,\,\,\,\,\,\,\,\, + {{\bf{E}}_h}\left\{ {{\bf{H}}_{r,h}^H{{\bf{\Theta }}_{vh}}{{\bf{G}}_v} + {\bf{H}}_{r,h}^H{{\bf{\Theta }}_{vh}}{{\bf{G}}_h} + {\bf{H}}_{d,h}^H} \right\}} \Bigr]{\bf{Fs}} + {\bf{Wz}},\\
\,\,\,\, = {\bf{W}}\Bigr[ {{{\bf{E}}_v}\left\{ {{\bf{H}}_{r,v}^H{{\bf{\Theta }}_{hv}}{\bf{G}} + {\bf{H}}_{d,v}^H} \right\} + {{\bf{E}}_h}\left\{ {{\bf{H}}_{r,h}^H{{\bf{\Theta }}_{vh}}{{\bf{G}}_{}} + {\bf{H}}_{d,h}^H} \right\}} \Bigr]{\bf{Fs}} + {\bf{Wz}},
\end{array}
\end{equation}
\textcolor{black}{ where \({\bf{G}} = {\bf{G}}_v^{} + {\bf{G}}_h^{}\).}

\textcolor{black}{ It is pertinent to observe that although we do not specifically optimize \(\left( {{{\bf{\Theta }}_{vv}},{{\bf{\Theta }}_{hh}}} \right)\) and \(\left( {{{\bf{U }}_{v}},{{\bf{U }}_{h}}} \right)\) in this work, our proposed algorithms can be readily extended to accommodate them. In addition, the energy normalization factor in the signal model is neglected. However, the incorporation of a factor of \(\frac{1}{{\sqrt 2 }}\) in \eqref{eq1} would resolve the normalization issue. It is also noteworthy that our simplifications result in a decline in the PD and combining gains in the system. Consequently, the gains of the DP-IRS over S-IRS presented in Section V can be regarded as the minimum gains. Further details on the gains of the DP-IRS over S-IRS are provided in Section V.}

\textcolor{black}{Let \({\mathbf{H}} = {{\mathbf{E}}_v}\left\{ {{\mathbf{H}}_{r,v}^H{{\mathbf{\Theta }}_v}{{\mathbf{G}}} + {\mathbf{H}}_{d,v}^H} \right\} + {{\mathbf{E}}_h}\left\{ {{\mathbf{H}}_{r,h}^H{{\mathbf{\Theta }}_h}{{\mathbf{G}}} + {\mathbf{H}}_{d,h}^H} \right\}\) denote a composite MIMO channel, where \({{\bf{\Theta }}_{v}}\) and \({{\bf{\Theta }}_{h}}\) represent \({{\bf{\Theta }}_{hv}}\) and \({{\bf{\Theta }}_{vh}}\), respectively, for notational brevity.} Then, the spectral efficiency of the system is given by
\begin{equation} \label{eq5}
c = {\log _2}\det ({{\mathbf{I}}_{}} + \frac{1}{{\sigma ^2}}{{\mathbf{W}}^\dag }{\mathbf{HF}}{{\mathbf{F}}^H}{{\mathbf{H}}^H}{\mathbf{W}})\,\,\,{\rm{bps/Hz}},
\end{equation}
where \(\bf{I}\) denotes an identity matrix of an appropriate size and \({{\sigma ^2}}\) is the noise variance. Correspondingly, the capacity maximization problem can be formulated as:
\begin{equation} \tag{P1} \label{eq6}
\begin{array}{l}
\mathop {\max }\limits_{{\bf{W}},{\bf{F}},{{\bf{E}}_v},{{\bf{E}}_h},{{\bf{\Theta }}_v},{{\bf{\Theta }}_h}} {\log _2}\det ({{\bf{I}}} + \frac{1}{{\sigma ^2}}{{\bf{W}}^\dag }{\bf{HF}}{{\bf{F}}^H}{{\bf{H}}^H}{\bf{W}})\\
{\rm{s}}{\rm{.t}}{\rm{. }}\,\,\,\,\,\,\,\,\,{\bf{H}} = {{\bf{E}}_v}\left\{ {{\bf{H}}_{r,v}^H{{\bf{\Theta }}_v}{{\bf{G}}} + {\bf{H}}_{d,v}^H} \right\} + {{\bf{E}}_h}\left\{ {{\bf{H}}_{r,h}^H{{\bf{\Theta }}_h}{{\bf{G}}} + {\bf{H}}_{d,h}^H} \right\},\\
\,\,\,\,\,\,\,\,\,\,\,\,\,\,\,\,{{\bf{\Theta }}_p} = {\rm{diag}}({e^{ - j{\phi _{p,1}}}},{e^{ - j{\phi _{p,2}}}},{e^{ - j{\phi _{p,3}}}}, \cdots ,{e^{ - j{\phi _{p,N}}}}),\,p \in \left\{ {v,h} \right\},\\
\,\,\,\,\,\,\,\,\,\,\,\,\,\,\,\,{{\bf{E}}_p} = {\rm{diag}}({e^{ - j{\gamma _{p,1}}}},{e^{ - j{\gamma _{p,2}}}},{e^{ - j{\gamma _{p,3}}}}, \cdots ,{e^{ - j{\gamma _{p,{N_r}}}}}),\,p \in \left\{ {v,h} \right\},\\
\,\,\,\,\,\,\,\,\,\,\,\,\,\,\,{\left\| {\bf{F}} \right\|^2} = {\frac{{{P_t}}}{2}}.
\end{array}
\end{equation}
The objective function of \eqref{eq6} can be demonstrated to be a non-concave over \({{\bf{E}}_v}\), \({{\bf{E}}_h}\), \({{\bf{\Theta }}_v}\), and \({{\bf{\Theta }}_h}\). Moreover, the unimodular restrictions in the second and third constraints of \eqref{eq6} are nonconvex. In general, there is no standard method for solving \eqref{eq6} optimally. In the following two sections, we propose three AO-based algorithms to efficiently solve this problem.

\section{Proposed Solution to Problem (P1)}
\label{Sect:ProposedWork}
Here, we describe the framework of the proposed algorithm. Specifically, we first divide the problem \eqref{eq6} into several non-convex subproblems, which are solved separately. Then, we propose the AO algorithm, which solves the subproblems alternately until the objective function converges. 

\subsection{Optimal Precoder, Combiner, and Number of Data Streams}
Here, we reformulate the problem (P1) to find the optimal precoder, combiner, and number of data streams for the system. For formulation, we assume \({{\bf{E}}_v} = {{\bf{E}}_h} = {{\bf{\Theta }}_v} = {{\bf{\Theta }}_h} = \bf{I}\). Correspondingly, the problem \eqref{eq6} is written as
\begin{equation} \tag{P1.1} \label{eq7}
\begin{array}{l}
\mathop {\max }\limits_{{\bf{W}},{\bf{F}}}\,\,\,\,\, {\log _2}\det ({{\bf{I}}} + \frac{1}{{\sigma ^2}}{{\bf{W}}^\dag }{{\bf{H}}_I}{\bf{F}}{{\bf{F}}^H}{\bf{H}}_I^H{\bf{W}})\\
{\rm{s}}{\rm{.t}}{\rm{. }}\,\,\,\,\,\,\,\,\,{{\bf{H}}_I} = {{\bf{H}}_{r,v}^H{{\bf{G}}} + {\bf{H}}_{d,v}^H}+ {{\bf{H}}_{r,h}^H{{\bf{G}}} + {\bf{H}}_{d,h}^H}, \\
\,\,\,\,\,\,\,\,\,\,\,\,\,\,\,{\left\| {\bf{F}} \right\|^2} = {\frac{{{P_t}}}{2}}.
\end{array}
\end{equation}
Problem \eqref{eq7} is a conventional MIMO capacity maximization problem that can be solved by singular value decomposition (SVD) and the water-filling power allocation technique \cite{tse2005fundamentals}. Let \({\rm{svd}}\left( {{{\bf{H}}_I}} \right) = {\bf{U\Sigma }}{{\bf{V}}^H}\), where \({\mathbf{V}} \in {\mathbb{C}^{{N_t} \times R}}\) with \({\rm{rank}}\left( {{{\bf{H}}_I}} \right) \le \min \left( {{N_t},{N_r}} \right)\) denotes the maximum number of data streams that can be transferred over \({{\mathbf{H}}_I}\). The optimal precoder \({\mathbf{F}}\) and combiner \({\mathbf{W}}\) can be written as:
\({\mathbf{F}} = {\mathbf{V}}{\mathbf{\Gamma} ^{\frac{1}{2}}}\)
and
\({\mathbf{W}} = {\mathbf{U}}\), respectively, where \({\mathbf{\Gamma }} = {\text{diag(}}{\rho _1},{\rho _2}, \cdots ,{\rho _{{N_s}}}{\text{)}}\) is an \({N_s} \times {N_s}\) diagonal matrix with \({\rho _{{n_s}}}\) being the transmit power allocated to the \({{n_s}}\)th data stream, \({n_s} = 1,2, \cdots ,{N_s}\), which is determined by the water-filling power allocation such that \(\sum\limits_{{n_s} = 1}^{{N_s}} {{\rho _{{n_s}}} \le } {\frac{{{P_t}}}{2}}\). By inserting \({\bf{F}}\) and \({\bf{W}}\) into \eqref{eq6}, the problem for \({{{\bf{\Theta }}_v}}\), \({{{\bf{\Theta }}_h}}\), \({{{\bf{E }}_v}}\), and \({{{\bf{E }}_h}}\) can be expressed as
\begin{equation} \tag{P1.2} \label{eq8}
\begin{array}{l}
\mathop {\max }\limits_{{{\bf{\Theta }}_v},{{\bf{\Theta }}_h},{{\bf{E}}_v},{{\bf{E}}_h}} \,\,\,\,\sum\limits_{{n_s} = 1}^{{N_s}} {{{\log }_2}\left( {1 + \frac{1}{{{\sigma ^2}}}{{\left| {{\bf{w}}_{{n_s}}^*\left\{ {{{\bf{E}}_v}\left( {{\bf{H}}_{r,v}^H{{\bf{\Theta }}_v}{{\bf{G}}} + {\bf{H}}_{d,v}^H} \right) + {{\bf{E}}_h}\left( {{\bf{H}}_{r,h}^H{{\bf{\Theta }}_h}{{\bf{G}}} + {\bf{H}}_{d,h}^H} \right)} \right\}{\bf{f}}_{{n_s}}^{}} \right|}^2}} \right)} \\
{\rm{s}}{\rm{.t}}{\rm{. }}\,\,\,\,\,\,\,\,\,{{\bf{\Theta }}_p} = {\rm{diag}}({e^{ - j{\phi _{p,1}}}},{e^{ - j{\phi _{p,2}}}},{e^{ - j{\phi _{p,3}}}}, \cdots ,{e^{ - j{\phi _{p,N}}}}),\,p \in \left\{ {v,h} \right\},\\
\,\,\,\,\,\,\,\,\,\,\,\,\,\,\,\,{{\bf{E}}_p} = {\rm{diag}}({e^{ - j{\gamma _{p,1}}}},{e^{ - j{\gamma _{p,2}}}},{e^{ - j{\gamma _{p,3}}}}, \cdots ,{e^{ - j{\gamma _{p,{N_r}}}}}),\,p \in \left\{ {v,h} \right\},\\
\,\,\,\,\,\,\,\,\,\,\,\,\,\,\,{\left\| {{{\bf{f}}_{{n_s}}}} \right\|^2} = {\frac{{{\rho _{{n_s}}}}}{2}},\,{n_s} = 1,2, \cdots ,{N_s},
\end{array}
\end{equation}
where the objective function is the summation of the spectral efficiencies provided by \({N_s}\) data streams \cite{9777870}. The problem \eqref{eq8} is a non-convex optimization problem with a non-concave objective function.
In the next two subsections, we solve \eqref{eq8} for \(\left( {{{\bf{\Theta }}_v},{{\bf{\Theta }}_h}} \right)\) and \(\left( {{{\bf{E}}_v},{{\bf{E}}_h}} \right)\).

\subsection{DP-IRS Phases Optimization}
By assuming \({{\bf{E}}_v} = {{\bf{E}}_h} = {\bf{I}}\), the problem \eqref{eq8} can be written as
\begin{equation} \tag{P1.3} \label{eq9}
\begin{array}{l}
\mathop {\max }\limits_{{{\bf{\Theta }}_v},{{\bf{\Theta }}_h}} \,\,\,\,\sum\limits_{{n_s} = 1}^{{N_s}} {{{\log }_2}\left( {1 + \frac{1}{{{\sigma ^2}}}{{\left| {{\bf{w}}_{{n_s}}^*\left\{ { {{\bf{H}}_{r,v}^H{{\bf{\Theta }}_v}{{\bf{G}}} + {\bf{H}}_{d,v}^H}+{{\bf{H}}_{r,h}^H{{\bf{\Theta }}_h}{{\bf{G}}} + {\bf{H}}_{d,h}^H} } \right\}{\bf{f}}_{{n_s}}^{}} \right|}^2}} \right)} \\
{\rm{s}}{\rm{.t}}{\rm{. }}\,\,\,\,\,\,\,\,\,{{\bf{\Theta }}_p} = {\rm{diag}}({e^{ - j{\phi _{p,1}}}},{e^{ - j{\phi _{p,2}}}},{e^{ - j{\phi _{p,3}}}}, \cdots ,{e^{ - j{\phi _{p,N}}}}),\,p \in \left\{ {v,h} \right\},\\
\,\,\,\,\,\,\,\,\,\,\,\,\,\,\,{\left\| {{{\bf{f}}_{{n_s}}}} \right\|^2} = {\rho _{{n_s}}},\,{n_s} = 1,2, \cdots ,{N_s}.
\end{array}
\end{equation}
By the change of variables \({\bf{w}}_{{n_s}}^*{\bf{H}}_{d,p}^H{\bf{f}}_{{n_s}}^{} = \beta _{p,{n_s}}^{}\), \({\bf{w}}_{{n_s}}^*{\bf{H}}_{r,p}^H{{\bf{\Theta }}_p}{{\bf{G}}}{\bf{f}}_{{n_s}}^{} = {\bf{l}}_p^H{{\boldsymbol{\alpha}} _{p,{n_s}}}\), where \({\bf{l}}_p^H = \left[ {{e^{ - j{\phi _{p,1}}}},{e^{ - j{\phi _{p,2}}}},{e^{ - j{\phi _{p,3}}}}, \cdots ,{e^{ - j{\phi _{p,N}}}}} \right]\,\), \({{\boldsymbol{\alpha}} _{p,{n_s}}} = {\rm{diag}}\left( {{\bf{w}}_{{n_s}}^*{\bf{H}}_{r,p}^H} \right){{\bf{G}}}{\bf{f}}_{{n_s}}^{}\), and \(p \in \left\{ {v,h} \right\}\), \eqref{eq9} is rewritten as
\begin{equation} \tag{P1.4} \label{eq11}
\begin{array}{l}
\mathop {\max }\limits_{{{\bf{l}}_v},{{\bf{l}}_h}} \,\,\,\,\sum\limits_{{n_s} = 1}^{{N_s}} {{{\log }_2}\left( {1 + \frac{1}{{{\sigma ^2}}}{{\left| {{\bf{l}}_v^H{{\boldsymbol{\alpha}} _{v,{n_s}}}\, + {\bf{l}}_h^H{{\boldsymbol{\alpha}} _{h,{n_s}}} + \beta _{v,{n_s}}^{} + \beta _{h,{n_s}}^{}} \right|}^2}} \right)} \\
{\rm{s}}{\rm{.t}}{\rm{. }}\,\,\,\,\,\,\,\,\,\left| {{{\bf{l}}_{n,p}}} \right| = 1,\,\,\,\,n = 1, \cdots ,N,\,\,\,\,p \in \left\{ {v,h} \right\}.
\end{array}
\end{equation}
 Next, we solve the problems \eqref{eq11} for \({{{\bf{l}}_v}}\) and \({{{\bf{l}}_h}}\), separately.

\subsubsection{Optimization of Vertical Phase Shifts}

Here, we solve \eqref{eq11} for \({{{\bf{l}}_v}}\). Let \({\bf{l}}_h^{} = \,\left[ {1,1, \cdots ,1} \right]\) and \({\bf{l}}_h^H{{\boldsymbol{\alpha}} _{h,{n_s}}} + \beta _{v,{n_s}}^{} + \beta _{h,{n_s}}^{} = {\eta _{v,{n_s}}}\). \eqref{eq11} can be written as
\begin{equation} \tag{P1.5} \label{eq12}
\begin{array}{l}
\mathop {\max }\limits_{{{\bf{l}}_v}} \,\,\,\,\sum\limits_{{n_s} = 1}^{{N_s}} {{{\log }_2}\left( {1 + \frac{1}{{{\sigma ^2}}}{{\left| {{\bf{l}}_v^H{{\boldsymbol{\alpha}} _{v,{n_s}}}\, + {\eta _{v,{n_s}}}} \right|}^2}} \right)} \\
{\rm{s}}{\rm{.t}}{\rm{. }}\,\,\,\,\,\,\,\,\,\left| {{{\bf{l}}_{n,v}}} \right| = 1,\,\,\,\,n = 1, \cdots ,N.\,
\end{array}
\end{equation}
Despite being simplified, problem \eqref{eq12} remains a nonconvex optimization problem with a nonconcave objective function and nonconvex unit modulus constraint. Nevertheless, we convert \eqref{eq12} into a semi-definite program using the SDR method, which can be efficiently performed by a CVX solver \cite{cvx}. Inserting \({\left| {{\bf{l}}_v^H{{\boldsymbol{\alpha}} _{v,{n_s}}}\, + {\eta _{v,{n_s}}}} \right|^2} = {\bf{l}}_v^H{{\boldsymbol{\alpha}} _{v,{n_s}}}{\boldsymbol{\alpha}} _{v,{n_s}}^H{\bf{l}}_v^{} + {\bf{l}}_v^H{{\boldsymbol{\alpha}} _{v,{n_s}}}\eta _{v,{n_s}}^H + \eta _{v,{n_s}}^{}{\boldsymbol{\alpha}} _{v,{n_s}}^H{\bf{l}}_v^{} + {\left| {{\eta _{v,{n_s}}}} \right|^2}\) in \eqref{eq12}, we have
\begin{equation} \tag{P1.6} \label{eq13}
\begin{array}{l}
\mathop {\max }\limits_{{{\bf{l}}_v}} \,\,\,\,\sum\limits_{{n_s} = 1}^{{N_s}} {{{\log }_2}\left( {1 + \frac{1}{{{\sigma ^2}}}\left\{ {{\bf{l}}_v^H{{\boldsymbol{\alpha}} _{v,{n_s}}}{\boldsymbol{\alpha}} _{v,{n_s}}^H{\bf{l}}_v^{} + {\bf{l}}_v^H{{\boldsymbol{\alpha}} _{v,{n_s}}}\eta _{v,{n_s}}^H + \eta _{v,{n_s}}^{}{\boldsymbol{\alpha}} _{v,{n_s}}^H{\bf{l}}_v^{} + {{\left| {{\eta _{v,{n_s}}}} \right|}^2}} \right\}} \right)} \\
{\rm{s}}{\rm{.t}}{\rm{. }}\,\,\,\,\,\,\,\,\,\left| {{{\bf{l}}_{n,v}}} \right| = 1,\,\,\,\,n = 1, \cdots ,N.\,
\end{array}
\end{equation}
By defining \({\bf{O}}_{v,{n_s}}^{} = \left[ {\begin{array}{*{20}{c}}
{{{\boldsymbol{\alpha}} _{v,{n_s}}}{\boldsymbol{\alpha}} _{v,{n_s}}^H}&{{{\boldsymbol{\alpha}} _{v,{n_s}}}\eta _{v,{n_s}}^H}\\
{{\boldsymbol{\alpha}} _{v,{n_s}}^H\eta _{v,{n_s}}^H}&0
\end{array}} \right]\) and \(\overline {{{\bf{l}}_v}}= \left[ {\begin{array}{*{20}{c}}
{{{\bf{l}}_v}}\\.
\psi 
\end{array}} \right]\), where \(\psi \) is an auxiliary variable, \eqref{eq13} can be homogenized \cite{wu2019beamforming} as
\begin{equation} \tag{P1.7} \label{eq14}
\begin{array}{l}
\mathop {\max }\limits_{\overline {{{\bf{l}}_v}} } \,\,\,\,\sum\limits_{{n_s} = 1}^{{N_s}} {{{\log }_2}\bigg( {1 + \frac{1}{{{\sigma ^2}}}\big\{ {\overline {{\bf{l}}_h^{}} ^H} {\bf{O}}_{v,{n_s}}^{}\overline {{{\bf{l}}_v}}+ {{\left| {{\eta _{v,{n_s}}}} \right|}^2}} \big\}} \bigg) \\
{\rm{s}}{\rm{.t}}{\rm{. }}\,\,\,\,\,\,\,\,\,\left| {\overline {{{\bf{l}}_{n,v}}} } \right| = 1,\,\,\,\,n = 1, \cdots ,N + 1.\,
\end{array}
\end{equation}
In \eqref{eq14}, we insert \({\overline {{\bf{l}}_h^{}} ^H} {\bf{O}}_{v,{n_s}}^{}\overline {{{\bf{l}}_v}}= {\rm{trace}}\left( {{\bf{O}}_{v,{n_s}}^{}{\bf{L}}_v^{}} \right)\), where \(\,{\bf{L}}_v^{} = {\overline {{{\bf{l}}_v}}}\,{\overline {{\bf{l}}_h^{}} ^H} \) must satisfy the constraints \({\bf{L}}_v^{}\underline  \succ  0\) and \({\rm{rank}}\left( {{\bf{L}}_v^{}} \right) = 1\), we have
\begin{equation} \tag{P1.8} \label{eq15}
\begin{array}{l}
\,\mathop {\max }\limits_{{\bf{L}}_v^{}} \,\,\,\,\sum\limits_{{n_s} = 1}^{{N_s}} {{{\log }_2}\Bigr( {1 + \frac{1}{{{\sigma ^2}}}\left( {{\rm{trace}}\left( {{\bf{O}}_{v,{n_s}}^{}{\bf{L}}_v^{}} \right) + {{\left| {{\eta _{v,{n_s}}}} \right|}^2}} \right)} \Bigr)} \\
{\rm{s}}{\rm{.t}}{\rm{. }}\,\,\,\,\,\,\,\,\,{\bf{L}}_{\left( {n,n} \right),v}^{} = 1,\,\,\,\,n = 1, \cdots ,N + 1,\,\\
\,\,\,\,\,\,\,\,\,\,\,\,\,\,\,\,{\bf{L}}_v^{}\underline  \succ  0,\\
\,\,\,\,\,\,\,\,\,\,\,\,\,\,\,{\rm{rank}}\left( {{\bf{L}}_v^{}} \right) = 1.
\end{array}
\end{equation}
Although the objective function is concave, the problem \eqref{eq15} remains a non-convex optimization problem because of the non-convex rank-one constraint, i.e., \({\rm{rank}}\left( {{\bf{L}}_v^{}} \right) = 1\). Relaxing this constraint, \eqref{eq15} is given by
\begin{equation} \tag{P1.9} \label{eq16}
\begin{array}{l}
\mathop {\max }\limits_{{\bf{L}}_v^{}} \,\,\,\,\sum\limits_{{n_s} = 1}^{{N_s}} {{{\log }_2}\Bigr( {1 + \frac{1}{{{\sigma ^2}}}\left( {{\rm{trace}}\left( {{\bf{O}}_{v,{n_s}}^{}{\bf{L}}_v^{}} \right) + {{\left| {{\eta _{v,{n_s}}}} \right|}^2}} \right)} \Bigr)} \\
{\rm{s}}{\rm{.t}}{\rm{. }}\,\,\,\,\,\,\,\,\,{\bf{L}}_{\left( {n,n} \right),v}^{} = 1,\,\,\,\,n = 1, \cdots ,N + 1,\,\\
\,\,\,\,\,\,\,\,\,\,\,\,\,\,\,\,{\bf{L}}_v^{}\underline  \succ  0.
\end{array}
\end{equation}
The problem \eqref{eq16} is a convex optimization problem that can be solved using a CVX solver \cite{cvx}; however, two issues remain with its formulation.
\begin{enumerate}
    \item Solving \eqref{eq16} does not guarantee that the objective value, i.e.,
    
    \(\sum\limits_{{n_s} = 1}^{{N_s}} {{{\log }_2}\Bigr( {1 + \frac{1}{{{\sigma ^2}}}\left( {{\rm{trace}}\left( {{\bf{O}}_{v,{n_s}}^{}{\bf{L}}_v^{}} \right) + {{\left| {{\eta _{v,{n_s}}}} \right|}^2}} \right)} \Bigr)} \), monotonically increases in each iteration of the AO algorithm, which may result in the non-convergence behaviors of the algorithm.

    \item Because of the objective function and last constraint, i.e., log-sum expression in \eqref{eq16}, which belongs to the functions of the exponential family, using a CVX solver to solve \eqref{eq16} is excessively slow. More details on this issue are presented in \cite{cvxProblem}.
\end{enumerate}
To address the first issue, we introduce another constraint and update \eqref{eq16} as follows:
\begin{equation} \tag{P1.10} \label{eq17a}
\begin{array}{l}
\,\mathop {\max }\limits_{{\bf{L}}_v^{}} \,\,\,\,\sum\limits_{{n_s} = 1}^{{N_s}} {{{\log }_2}\Bigr( {1 + \frac{1}{{{\sigma ^2}}}\left( {{\rm{trace}}\left( {{\bf{O}}_{v,{n_s}}^{}{\bf{L}}_v^{}} \right) + {{\left| {{\eta _{v,{n_s}}}} \right|}^2}} \right)} \Bigr)} \\
{\rm{s}}{\rm{.t}}{\rm{. }}\,\,\,\,\,\,\,\,\,{{\bf{L}}_{{v_{\left( {n,n} \right)}}}} = 1,\,\,\,\,n = 1, \cdots ,N + 1,\,\\
\,\,\,\,\,\,\,\,\,\,\,\,\,\,\,\,{\bf{L}}_v^{}\underline  \succ  0,\\
\,\,\,\,\,\,\,\,\,\,\,\,\,\,\,\sum\limits_{{n_s} = 1}^{{N_s}} {{{\log }_2}\Bigr( {1 + \frac{1}{{{\sigma ^2}}}\left( {{\rm{trace}}\left( {{\bf{O}}_{v,{n_s}}^{}{\bf{L}}_v^{}} \right) + {{\left| {{\eta _{v,{n_s}}}} \right|}^2}} \right)} \Bigr)}  \ge \lambda _{{{\bf{L}}_v}}^{}, 
\end{array}
\end{equation}
where \(\lambda _{{{\bf{L}}_v}}^{}\) represents a positive real number chosen in each iteration of AO, such that the objective value remains non-decreasing and guarantees convergence. More details on the selection of \(\lambda _{{{\bf{L}}_v}}^{}\) are provided in Section IV-D.

To address the second issue, the log-sum expressions in \eqref{eq17a} must be discarded. In \cite{cvxProblem}, it is demonstrated that an optimization problem involving the log-sum expression can be replaced with the geometric mean. Specifically, let\\\({{\bf{x}}_{{\bf{L}}_v}} = \Bigr[ {\Bigr( {1 + \frac{1}{{{\sigma ^2}}}\left( {{\rm{trace}}\left( {{\bf{O}}_{v,1}^{}{\bf{L}}_v^{}} \right) + {{\left| {{\eta _{v,1}}} \right|}^2}} \right)} \Bigr), \cdots ,\Bigr( {1 + \frac{1}{{{\sigma ^2}}}\left( {{\rm{trace}}\left( {{\bf{O}}_{v,{n_s}}^{}{\bf{L}}_v^{}} \right) + {{\left| {{\eta _{v,{n_s}}}} \right|}^2}} \right)} \Bigr)} \Bigr]\) be a vector containing the \(1+{\rm{SNR}}\) value of each data stream, then the objective function \({\rm{logsum}}\left( {\bf{x}} \right)\) and constraint \({\rm{logsum}}\left( {\bf{x}} \right)\, \ge \lambda _{{{\bf{L}}_v}}^{} \) in \eqref{eq17a} can be replaced by \({\rm{GM}}\left( {\bf{x}} \right)\) and \({\rm{GM}}\left( {\bf{x}} \right)\,\, \ge {\left( {\log \left( \lambda _{{{\bf{L}}_v}}^{}  \right)\,} \right)^{\frac{1}{{{\rm{length}}\left( {\bf{x}} \right)}}}} = \,{\left( {\log \left( \lambda _{{{\bf{L}}_v}}^{}  \right)\,} \right)^{\frac{1}{{{N_s}}}}} = \,{\lambda _{{{\bf{L}}_v}}^'}\), respectively. Consequently, we have
\begin{equation} \tag{P1.11} \label{eq17}
\begin{array}{l}
\mathop {\max }\limits_{{\bf{L}}_v^{}} \,\,\,\,\,\,{\rm{GM}}\left( {{{\bf{x}}_{{\bf{L}}_v}}} \right)\,\,\,\\
{\rm{s}}{\rm{.t}}{\rm{. }}\,\,\,\,\,\,\,\,\,\,\,{{\bf{x}}_{{\bf{L}}_v}} = \Bigr[ {\Bigr( {1 + \frac{1}{{{\sigma ^2}}}\left( {{\rm{trace}}\left( {{\bf{O}}_{v,1}^{}{\bf{L}}_v^{}} \right) + {{\left| {{\eta _{v,1}}} \right|}^2}} \right)} \Bigr), \cdots ,\Bigr( {1 + \frac{1}{{{\sigma ^2}}}\left( {{\rm{trace}}\left( {{\bf{O}}_{v,{n_s}}^{}{\bf{L}}_v^{}} \right) + {{\left| {{\eta _{v,{n_s}}}} \right|}^2}} \right)} \Bigr)} \Bigr]\\
\,\,\,\,\,\,\,\,\,\,\,\,\,\,\,\,\,\,{{\bf{L}}_{{v_{\left( {n,n} \right)}}}} = 1,\,\,\,\,n = 1, \cdots ,N + 1,\,\\
\,\,\,\,\,\,\,\,\,\,\,\,\,\,\,\,\,\,{\bf{L}}_v^{}\underline  \succ  0,\\
\,\,\,\,\,\,\,\,\,\,\,\,\,\,\,\,\,{\rm{GM}}\left( {{{\bf{x}}_{{\bf{L}}_v}}} \right) \ge {\lambda _{{{\bf{L}}_v}}^{'}}.
\end{array}
\end{equation}
Note that \eqref{eq17} can be written in a simpler form, which is a feasibility check problem without any objective function, as 
\begin{equation} \tag{P1.12} \label{eq18}
\begin{array}{l}
\mathop {{\rm{find}}}\limits_{} \,\,\,\,\,\,{\bf{L}}_v^{}\,\,\,\\
{\rm{s}}{\rm{.t}}{\rm{. }}\,\,\,\,\,\,\,\,\,\,\,{{\bf{x}}_{{\bf{L}}_v}} = \Bigr[ {\Bigr( {1 + \frac{1}{{{\sigma ^2}}}\left( {{\rm{trace}}\left( {{\bf{O}}_{v,1}^{}{\bf{L}}_v^{}} \right) + {{\left| {{\eta _{v,1}}} \right|}^2}} \right)} \Bigr), \cdots ,\Bigr( {1 + \frac{1}{{{\sigma ^2}}}\left( {{\rm{trace}}\left( {{\bf{O}}_{v,{n_s}}^{}{\bf{L}}_v^{}} \right) + {{\left| {{\eta _{v,{n_s}}}} \right|}^2}} \right)} \Bigr)} \Bigr]\\
\,\,\,\,\,\,\,\,\,\,\,\,\,\,\,\,\,\,{{\bf{L}}_{{v_{\left( {n,n} \right)}}}} = 1,\,\,\,\,n = 1, \cdots ,N + 1,\,\\
\,\,\,\,\,\,\,\,\,\,\,\,\,\,\,\,\,\,{\bf{L}}_v^{}\underline  \succ  0,\\
\,\,\,\,\,\,\,\,\,\,\,\,\,\,\,\,\,{\rm{GM}}\left( {{{\bf{x}}_{{\bf{L}}_v}}} \right) \ge {\lambda _{{{\bf{L}}_v}}^{'}}.
\end{array}
\end{equation}
However, because there is no objective function, solving \eqref{eq18} requires more iterations than solving \eqref{eq17} for the convergence of the AO algorithm, as will be demonstrated in Section VI. To accumulate the benefits of \eqref{eq17} and \eqref{eq18}, i.e., simpler formulation and fewer iterations, into one formulation, we introduce a slack variable \(\xi _{{{\bf{L}}_v}}^{}\) and reformulate the problem \eqref{eq18} as
\begin{equation} \tag{P1.13} \label{eq19} 
\begin{array}{l}
\mathop {\max }\limits_{{\bf{L}}_v^{}} \,\,\,\,\xi _{{{\bf{L}}_v}}^{} \,\,\,\\
{\rm{s}}{\rm{.t}}{\rm{. }}\,\,\,\,\,\,\,\,\,\,\,{{\bf{x}}_{{\bf{L}}_v}} = \Bigr[ {\Bigr( {1 + \frac{1}{{{\sigma ^2}}}\left( {{\rm{trace}}\left( {{\bf{O}}_{v,1}^{}{\bf{L}}_v^{}} \right) + {{\left| {{\eta _{v,1}}} \right|}^2}} \right)} \Bigr), \cdots ,\Bigr( {1 + \frac{1}{{{\sigma ^2}}}\left( {{\rm{trace}}\left( {{\bf{O}}_{v,{n_s}}^{}{\bf{L}}_v^{}} \right) + {{\left| {{\eta _{v,{n_s}}}} \right|}^2}} \right)} \Bigr)} \Bigr]\\
\,\,\,\,\,\,\,\,\,\,\,\,\,\,\,\,\,\,{{\bf{L}}_{{v_{\left( {n,n} \right)}}}} = 1,\,\,\,\,n = 1, \cdots ,N + 1,\,\\
\,\,\,\,\,\,\,\,\,\,\,\,\,\,\,\,\,\,{\bf{L}}_v^{}\underline  \succ  0,\\
\,\,\,\,\,\,\,\,\,\,\,\,\,\,\,\,\,{\rm{GM}}\left( {{{\bf{x}}_{{\bf{L}}_v}}} \right) \ge {\lambda _{{{\bf{L}}_v}}^'} + \xi _{{{\bf{L}}_v}}^{}. 
\end{array}
\end{equation}
The formulation of \eqref{eq19} is simpler than that of \eqref{eq17}, and solving it requires fewer iterations than \eqref{eq18}. More details regarding the convergence behaviors and time complexities of these formulations are provided in Section VI.
Solving \eqref{eq19} may result in a higher-rank solution; hence, retrieving \({{\bf{\Theta }}_v}\) from \({{\bf{L}}_v^{}}\) might require more steps, i.e., Gaussian randomization. Details are provided in \cite{wu2018intelligent}. 

\subsubsection{Optimization of Horizontal Phase Shifts}

Let \({\bf{l}}_v^H{{\boldsymbol{\alpha}} _{v,{n_s}}} + \beta _{v,{n_s}}^{} + \beta _{h,{n_s}}^{} = {\eta _{h,{n_s}}}\), where \({\bf{l}}_v^{}\) is retrieved from \({{\bf{L}}_v^{}}\) by solving \eqref{eq17}. Then the problem \eqref{eq11}  for \({\bf{l}}_h^{}\) can be written as
\begin{equation} \tag{P1.14} \label{eq20}
\begin{array}{l}
\mathop {\max }\limits_{{{\bf{l}}_h}} \,\,\,\,\sum\limits_{{n_s} = 1}^{{N_s}} {{{\log }_2}\left( {1 + \frac{1}{{{\sigma ^2}}}{{\left| {{\bf{l}}_h^H{{\boldsymbol{\alpha}} _{v,{n_s}}}\, + {\eta _{h,{n_s}}}} \right|}^2}} \right)} \\
{\rm{s}}{\rm{.t}}{\rm{. }}\,\,\,\,\,\,\,\,\,\left| {{{\bf{l}}_{n,h}}} \right| = 1,\,\,\,\,n = 1, \cdots ,N.\,
\end{array}
\end{equation}
Following the steps similar to those for the vertical polarization phases, we obtain an equivalent formulation for \eqref{eq20} as
\begin{equation} \tag{P1.15} \label{eq21} {
\begin{array}{l}
\mathop {\max }\limits_{{\bf{L}}_h^{}} \,\,\,\,\xi _{{{\bf{L}}_h}}^{} \,\,\,\\
{\rm{s}}{\rm{.t}}{\rm{. }}\,\,\,\,\,\,\,\,\,\,\,{{\bf{x}}_{{\bf{L}}_h}} = \Bigr[ {\Bigr( {1 + \frac{1}{{{\sigma ^2}}}\left( {{\rm{trace}}\left( {{\bf{O}}_{h,1}^{}{\bf{L}}_h^{}} \right) + {{\left| {{\eta _{h,1}}} \right|}^2}} \right)} \Bigr), \cdots ,\Bigr( {1 + \frac{1}{{{\sigma ^2}}}\left( {{\rm{trace}}\left( {{\bf{O}}_{h,{n_s}}^{}{\bf{L}}_h^{}} \right) + {{\left| {{\eta _{h,{n_s}}}} \right|}^2}} \right)} \Bigr)} \Bigr]\\
\,\,\,\,\,\,\,\,\,\,\,\,\,\,\,\,\,\,{{\bf{L}}_{{h_{\left( {n,n} \right)}}}} = 1,\,\,\,\,n = 1, \cdots ,N + 1,\,\\
\,\,\,\,\,\,\,\,\,\,\,\,\,\,\,\,\,\,{\bf{L}}_h^{}\underline  \succ  0,\\
\,\,\,\,\,\,\,\,\,\,\,\,\,\,\,\,\,{\rm{GM}}\left( {{{\bf{x}}_{{\bf{L}}_h}}} \right) \ge {\lambda _{{{\bf{L}}_h}}^'} + \xi _{{{\bf{L}}_h}}^{}, 
\end{array} }
\end{equation}
where \({\bf{O}}_{h,{n_s}}^{} = \left[ {\begin{array}{*{20}{c}}
{{{\boldsymbol{\alpha}} _{h,{n_s}}}{\boldsymbol{\alpha}} _{h,{n_s}}^H}&{{{\boldsymbol{\alpha}} _{h,{n_s}}}\eta _{h,{n_s}}^H}\\
{{\boldsymbol{\alpha}} _{h,{n_s}}^H\eta _{h,{n_s}}^H}&0
\end{array}} \right]\), \({\bf{L}}_h^{} = \overline {{{\bf{l}}_h}}\,{\overline {{{\bf{l}}_h}} ^H} \), \(\overline {{{\bf{l}}_h}}  = \left[ {\begin{array}{*{20}{c}}
{{{\bf{l}}_h}}\\
\psi 
\end{array}} \right]\,\), \(\psi \) is a auxiliary variable, \(\xi _{{{\bf{L}}_h}}^{}\) is a slack variable, and \(\lambda _{{{\bf{L}}_h}}^'\) is positive real number.

\subsection{Optimal Phase Shifters for Receive DP Antennas}

Given \({{\bf{\Theta }}_v}\) and \({{\bf{\Theta }}_h}\) from \eqref{eq19}  and \eqref{eq21} , respectively, the problem \eqref{eq8}  for \({{\bf{E}}_v}\) and \({{\bf{E}}_h}\) can be rewritten as
\begin{equation} \tag{P1.16} \label{eq22}
\begin{array}{l}
\mathop {\max }\limits_{{{\bf{E}}_v},{{\bf{E}}_h}} \,\,\,\,\sum\limits_{{n_s} = 1}^{{N_s}} {{{\log }_2}\left( {1 + \frac{1}{{{\sigma ^2}}}{{\left| {{\bf{w}}_{{n_s}}^*{{\bf{E}}_v}{\bf{h}}_{v,{n_s}}^{} + {\bf{w}}_{{n_s}}^*{{\bf{E}}_h}{\bf{h}}_{h,{n_s}}^{}} \right|}^2}} \right)} \\
{\rm{s}}{\rm{.t}}{\rm{. }}\,\,\,\,\,\,\,\,\,{{\bf{E}}_p} = {\rm{diag}}({e^{ - j{\gamma _{p,1}}}},{e^{ - j{\gamma _{p,2}}}},{e^{ - j{\gamma _{p,3}}}}, \cdots ,{e^{ - j{\gamma _{p,{N_r}}}}}),\,p \in \left\{ {v,h} \right\}\\
\,\,\,\,\,\,\,\,\,\,\,\,\,\,\,\,{\bf{h}}_{p,{n_s}}^{} = {\bf{H}}_{r,p}^H{{\bf{\Theta }}_p}{{\bf{G}}}{\bf{f}}_{{n_s}}^{} + {\bf{H}}_{d,p}^H{\bf{f}}_{{n_s}}^{},\,\,p \in \left\{ {v,h} \right\}\\
\,\,\,\,\,\,\,\,\,\,\,\,\,\,\,\,{\left\| {{{\bf{f}}_{{n_s}}}} \right\|^2} = {\rho _{{n_s}}},\,\,{n_s} = 1,2, \cdots ,{N_s}.\,
\end{array}
\end{equation}
By applying the change of variables \({\bf{w}}_{{n_s}}^*{{\bf{E}}_p}{\bf{h}}_{p,{n_s}}^{} = {\bf{e}}_p^{T}{\bf{h}}_{p,{n_s}}^'\), where \({\bf{h}}_{p,{n_s}}^' = {\rm{diag}}\left( {{\bf{w}}_{{n_s}}^*} \right)\cdot{\bf{h}}_{p,{n_s}}^{}\), \({\bf{e}}_p^{T} = \left[ {{e^{ - j{\gamma _{p,1}}}},{e^{ - j{\gamma _{p,2}}}},{e^{ - j{\gamma _{p,3}}}}, \cdots ,{e^{ - j{\gamma _{p,{N_r}}}}}} \right]\), and \(p \in \left\{ {v,h} \right\}\), \eqref{eq22} is reduced to
\begin{equation} \tag{P1.17} \label{eq23}
\begin{array}{l}
\mathop {\max }\limits_{{{\bf{e}}_v^{}},{{\bf{e}}_h}^{}} \,\,\,\,\sum\limits_{{n_s} = 1}^{{N_s}} {{{\log }_2}\left( {1 + \frac{1}{{{\sigma ^2}}}{{\left| {{\bf{e}}_v^{T}{\bf{h}}_{v,{n_s}}^' + {\bf{e}}_h^{T}{\bf{h}}_{h,{n_s}}^'} \right|}^2}} \right)} \\
{\rm{s}}{\rm{.t}}{\rm{. }}\,\,\,\,\,\,\,\,\,{\bf{e}}_p^{T} = \left[ {{e^{ - j{\gamma _{p,1}}}},{e^{ - j{\gamma _{p,2}}}},{e^{ - j{\gamma _{p,3}}}}, \cdots ,{e^{ - j{\gamma _{p,{N_r}}}}}} \right],\,p \in \left\{ {v,h} \right\}\\
\,\,\,\,\,\,\,\,\,\,\,\,\,\,\,\,{\bf{h}}_{p,{n_s}}^' = {\rm{diag}}\left( {{\bf{w}}_{{n_s}}^*} \right)\cdot{\bf{h}}_{p,{n_s}}^{},\,\,\,{n_s} = 1,2, \cdots ,{N_s}.\,
\end{array}
\end{equation}
Next, we solve \eqref{eq23} separately for \({{{\bf{e}}_v}}\) and \({{{\bf{e}}_h}}\).

\subsubsection{Receive Vertical Phase Shifts Optimization}
Here, we solve the problem \eqref{eq23} for the vertical phase shifters of the receive DP antennas. Let \({\bf{e}}_h^{T} = \left[ {1,1, \cdots ,1} \right]\) and \({\bf{e}}_h^{T}{\bf{h}}_{h,{n_s}}^' = {\delta _{h,{n_s}}}\), the problem \eqref{eq23} for \({{\bf{e}}_v^{}}\) can be written as
\begin{equation} \tag{P1.18} \label{eq24}
\begin{array}{l}
\mathop {\max }\limits_{{{\bf{e}}_v}} \,\,\,\,\sum\limits_{{n_s} = 1}^{{N_s}} {{{\log }_2}\left( {1 + \frac{1}{{{\sigma ^2}}}{{\left| {{\bf{e}}_v^{T}{\bf{h}}_{v,{n_s}}^' + {{\delta _{h,{n_s}}}}} \right|}^2}} \right)} \\
{\rm{s}}{\rm{.t}}{\rm{. }}\,\,\,\,\,\,\,\,\,{\bf{e}}_v^{T} = \left[ {{e^{ - j{\gamma _{v,1}}}},{e^{ - j{\gamma _{v,2}}}},{e^{ - j{\gamma _{v,3}}}}, \cdots ,{e^{ - j{\gamma _{v,{N_r}}}}}} \right],\\
\,\,\,\,\,\,\,\,\,\,\,\,\,\,\,\,{\bf{h}}_{v,{n_s}}^' = {\rm{diag}}\left( {{\bf{w}}_{{n_s}}^*} \right)\cdot{\bf{h}}_{v,{n_s}}^{},\,\,\,{n_s} = 1,2, \cdots ,{N_s}.\,
\end{array}
\end{equation}
Inserting \(
\left| \mathbf{e}_v^{T} \mathbf{h}_{v,{n_s}}' + \delta_{h,{n_s}} \right|^2 = \mathbf{e}_v^{T} \mathbf{h}_{v,{n_s}}' \mathbf{h}_{v,{n_s}}^{{{'{H}}}} \mathbf{e}_{v}^{*} + \mathbf{e}_v^{T} \mathbf{h}_{v,{n_s}}' \delta_{h,{n_s}}^{H} + \delta_{h,{n_s}} \mathbf{h}_{v,{n_s}}^{{{'H}}} \mathbf{e}_v^{*} + \left| \delta_{h,{n_s}} \right|^2
\)
 in \eqref{eq24}, we have
\begin{equation} \tag{P1.19} \label{eq25}
\begin{array}{l}
\mathop {\max }\limits_{{{\bf{e}}_v}} \,\,\,\,\sum\limits_{{n_s} = 1}^{{N_s}} {{{\log }_2}\left( {1 + \frac{1}{{{\sigma ^2}}}\left( {{\bf{e}}_v^{T}{\bf{h}}_{v,{n_s}}^'{\bf{h}}_{v,{n_s}}^{{'H}}{\bf{e}}_v^* + {\bf{e}}_v^{T}{\bf{h}}_{v,{n_s}}^'\delta _{h,{n_s}}^H + {\delta _{h,{n_s}}}{\bf{h}}_{v,{n_s}}^{{'H}}{\bf{e}}_v^* + {{\left| {{\delta _{h,{n_s}}}} \right|}^2}} \right)} \right)} \\
{\rm{s}}{\rm{.t}}{\rm{. }}\,\,\,\,\,\,\,\,\,{\bf{e}}_v^{T} = \left[ {{e^{ - j{\gamma _{v,1}}}},{e^{ - j{\gamma _{v,2}}}},{e^{ - j{\gamma _{v,3}}}}, \cdots ,{e^{ - j{\gamma _{v,{N_r}}}}}} \right],\\
\,\,\,\,\,\,\,\,\,\,\,\,\,\,\,\,{\bf{h}}_{v,{n_s}}^' = {\rm{diag}}\left( {{\bf{w}}_{{n_s}}^*} \right)\cdot{\bf{h}}_{v,{n_s}}^{},\,\,\,{n_s} = 1,2, \cdots ,{N_s}.\,
\end{array}
\end{equation}
Similar to \eqref{eq12}, by defining \({\bf{S}}_{{n_s}}^v\, = \left[ {\begin{array}{*{20}{c}}
{{\bf{h}}_{v,{n_s}}^'{\bf{h}}_{v,{n_s}}^{{'H}}}&{{\bf{h}}_{v,{n_s}}^'\delta _{h,{n_s}}^H}\\
{{\bf{h}}_{v,{n_s}}^{{'H}}{\delta _{h,{n_s}}}}&0
\end{array}} \right]\), \({\overline {{{\bf{e}}_v}} ^{^T}} = \left[ {\begin{array}{*{20}{c}}
{{\bf{e}}_v^{T}}&\psi, 
\end{array}} \right]\), \({{\bf{\Xi}} _v} = {\overline {{{\bf{e}}_v}} ^{^*}} {\overline {{{\bf{e}}_v}} ^{^T}} \), and \({\overline {{{\bf{e}}_v}} ^{^T}} {\bf{S}}_{{n_s}}^v{{\overline {{{\bf{e}}_v}} ^{^*}}} = {\rm{trace}}\left( {{\bf{S}}_{{n_s}}^v{{\bf{\Xi}} _v}} \right)\), the SDR formulation of \eqref{eq25} can be written as
\begin{equation} \tag{P1.20} \label{eq26}
\begin{array}{l}
\mathop {\max }\limits_{{{\bf{\Xi}} _v}} \,\,\,\,\sum\limits_{{n_s} = 1}^{{N_s}} {{{\log }_2}\left( {1 + \frac{1}{{{\sigma ^2}}}\left( {{\rm{trace}}\left( {{\bf{S}}_{{n_s}}^v{{\bf{\Xi}} _v}} \right) + {{\left| {{{\delta _{h,{n_s}}}}} \right|}^2}} \right)} \right)} \\
{\rm{s}}{\rm{.t}}{\rm{. }}\,\,\,\,\,\,\,\,\,\,{{\bf{\Xi }}_{{v_{\left( {{n_r},{n_r}} \right)}}}} = 1,\,\,n = 1, \cdots ,{N_r} + 1,\\
\,\,\,\,\,\,\,\,\,\,\,\,\,\,\,\,\,\,{{\bf{\Xi}} _v}\underline \succ 0.
\end{array}
\end{equation}
Problem \eqref{eq26} is a convex optimization problem that can be solved by CVX \cite{cvx}; however, similar to \eqref{eq16}, solving \eqref{eq26} in each iteration of the AO algorithm does not guarantee a non-decreasing objective value. Hence, an equivalent formulation of \eqref{eq26} is given by
\begin{equation} \tag{P1.21} \label{eq27} \begin{array}{l}
\mathop {\max }\limits_{{{\bf{\Xi}} _v}} \,\,\,\,\xi _{{{\bf{\Xi }}_v}}^{} \\
{\rm{s}}{\rm{.t}}{\rm{. }}\,\,\,\,\,\,\,\,\,{{\bf{x}}_{{{\bf{\Xi}} _v}}} = \left[ {\Bigr( {1 + \frac{1}{{{\sigma ^2}}}\left( {{\rm{trace}}\left( {{\bf{S}}_1^v{{\bf{\Xi}} _v}} \right) + {{\left| {{\delta _{h,{n_s}}}} \right|}^2}} \right)} \Bigr),\,\, \cdots ,\,\Bigr( {1 + \frac{1}{{{\sigma ^2}}}\left( {{\rm{trace}}\left( {{\bf{S}}_{{n_s}}^v{{\bf{\Xi}} _v}} \right) + {{\left| {{\delta _{h,{n_s}}}} \right|}^2}} \right)} \Bigr)} \right]\\
\,\,\,\,\,\,\,\,\,\,\,\,\,\,\,\,{{\bf{\Xi }}_{{v_{\left( {{n_r},{n_r}} \right)}}}} = 1,\,\,n = 1, \cdots ,{N_r} + 1,\\
\,\,\,\,\,\,\,\,\,\,\,\,\,\,\,\,{{\bf{\Xi}} _v}\underline \succ 0\\
\,\,\,\,\,\,\,\,\,\,\,\,\,\,\,\,{\rm{GM}}\left( {{\bf{x}}_{{{\bf{\Xi}} _v}}} \right) \ge \lambda _{{{\bf{\Xi }}_v}}^' + \xi _{{{\bf{\Xi }}_v}}^{}, 
\end{array}
\end{equation}
where \(\xi _{{{\bf{\Xi }}_v}}^{}\) and \(\lambda _{{{\bf{\Xi }}_v}}^'\) are a slack variable and real positive number, respectively. 

\subsubsection{Receive Horizontal Phase Shifts Optimization}
By letting \({\bf{e}}_v^{T}{\bf{h}}_{v,{n_s}}^' = {\delta _{v,{n_s}}}\), where \({\bf{e}}_v^{}\) is retrieved from \({{\bf{\Xi}} _v}\) by solving \eqref{eq27}, the problem \eqref{eq23} for \({\bf{e}}_h^{}\) can be expressed as
\begin{equation} \tag{P1.22} \label{eq28}
\begin{array}{l}
\mathop {\max }\limits_{{{\bf{e}}_h}} \,\,\,\,\sum\limits_{{n_s} = 1}^{{N_s}} {{{\log }_2}\left( {1 + \frac{1}{{{\sigma ^2}}}{{\left| {{\bf{e}}_h^{T}{\bf{h}}_{h,{n_s}}^' + {{\delta _{v,{n_s}}}}} \right|}^2}} \right)} \\
{\rm{s}}{\rm{.t}}{\rm{. }}\,\,\,\,\,\,\,\,\,{\bf{e}}_h^{T} = \left[ {{e^{ - j{\gamma _{h,1}}}},{e^{ - j{\gamma _{h,2}}}},{e^{ - j{\gamma _{h,3}}}}, \cdots ,{e^{ - j{\gamma _{v,{N_r}}}}}} \right],\\
\,\,\,\,\,\,\,\,\,\,\,\,\,\,\,\,{\bf{h}}_{h,{n_s}}^' = {\rm{diag}}\left( {{\bf{w}}_{{n_s}}^*} \right)\cdot{\bf{h}}_{h,{n_s}}^{},\,\,\,{n_s} = 1,2, \cdots ,{N_s}.\,
\end{array}
\end{equation}
Problem \eqref{eq28} is similar to \eqref{eq24}; hence, by following similar steps, an equivalent formulation of \eqref{eq28} can be obtained as
\begin{equation} \tag{P1.23} \label{eq29} {
\begin{array}{l}
\mathop {\max }\limits_{{{\bf{\Xi}} _h}} \,\,\,\,\xi _{{{\bf{\Xi }}_h}}^{} \\
{\rm{s}}{\rm{.t}}{\rm{. }}\,\,\,\,\,\,\,\,\,{{\bf{x}}_{{{\bf{\Xi}} _h}}} = \left[ {\Bigr( {1 + \frac{1}{{{\sigma ^2}}}\left( {{\rm{trace}}\left( {{\bf{S}}_1^h{{\bf{\Xi}} _h}} \right) + {{\left| {{\delta _{v,{n_s}}}} \right|}^2}} \right)} \Bigr),\,\, \cdots ,\,\Bigr( {1 + \frac{1}{{{\sigma ^2}}}\left( {{\rm{trace}}\left( {{\bf{S}}_{{n_s}}^h{{\bf{\Xi}} _h}} \right) + {{\left| {{\delta _{v,{n_s}}}} \right|}^2}} \right)} \Bigr)} \right]\\
\,\,\,\,\,\,\,\,\,\,\,\,\,\,\,\,{{\bf{\Xi }}_{{h_{\left( {{n_r},{n_r}} \right)}}}} = 1,\,\,n = 1, \cdots ,{N_r} + 1,\\
\,\,\,\,\,\,\,\,\,\,\,\,\,\,\,\,{{\bf{\Xi}} _h}\underline \succ 0\\
\,\,\,\,\,\,\,\,\,\,\,\,\,\,\,\,{\rm{GM}}\left( {\bf{x}} \right) \ge \lambda _{{{\bf{\Xi }}_h}}^' + \xi _{{{\bf{\Xi }}_h}}^{}, 
\end{array}}
\end{equation}
where \({\bf{S}}_{{n_s}}^h\, = \left[ {\begin{array}{*{20}{c}}
{{\bf{h}}_{h,{n_s}}^{'}{\bf{h}}_{h,{n_s}}^{{'{H}}}}&{{\bf{h}}_{h,{n_s}}^{'}{\delta} _{v,{n_s}}^{H}}\\
{{\bf{h}}_{h,{n_s}}^{{{'}^{H}}}{{\delta} _{v,{n_s}}}}&0
\end{array}} \right]\), \({{\bf{\Xi}} _h} = {\overline {{{\bf{e}}_h}} ^{*}} {\overline {{{\bf{e}}_h}} ^{T}} \), and \({\overline {{{\bf{e}}_h}} ^{T}} = \left[ {\begin{array}{*{20}{c}}
{{\bf{e}}_h^{T}}&\psi 
\end{array}} \right]\).

\subsection{Overall Algorithm}
The proposed AO algorithm solves \eqref{eq7}, \eqref{eq19}, \eqref{eq21}, \eqref{eq27}, and \eqref{eq29} alternately until the objective function of \eqref{eq6} converges. To maintain the monotonically non-decreasing behavior, the maximized objective value \(\xi \) in each problem is adopted as the \(\lambda ^'\) of the next problem. The overall algorithm is summarized in Algorithm \ref{alg:sdr}. 
\begin{algorithm}[H] \caption{SDR-based Alternating Optimization Algorithm}\label{alg:sdr}
\begin{algorithmic}[1]
\State {Set iteration \(k = 0\) and initialize \({{\bf{\Theta }}_v^k} = {{\bf{I}}_{{{\bf{\Theta }}_v}}}\), \({{\bf{\Theta }}_h^k} = {{\bf{I}}_{{{\bf{\Theta }}_h}}}\), \({{\bf{E }}_v^k} = {{\bf{I}}_{{{\bf{E }}_v}}}\), \({{\bf{E }}_h^k} = {{\bf{I}}_{{{\bf{E }}_h}}}\), and \(\xi _{{{\bf{\Xi }}_h}}^{k}=0\).}

\State {Solve \eqref{eq7} using the SVD and water-filling technique, and denote the optimized \(\bf{W}\), \({\bf{F}}\), and \({{N_s}}\) as \({{\bf{W}}^k}\), \({{\bf{F}}^k}\), and \(N_s^k\), respectively.}

\REPEAT

\State {Given \({\bf{E }}_v^{k}\), \({\bf{E }}_h^{k}\), and \({\bf{\Theta }}_h^k\), put \(\lambda _{{{\bf{L}}_v}}^{'k} = \xi _{{{\bf{\Xi }}_h}}^k\), solve \eqref{eq19}, retrieve optimal \({\bf{\Theta}}_v^{k+1}\) from \({{{\bf{L }}_v}}\) using Gaussian randomization, and denote the maximized objective value as \(\xi _{{{\bf{L }}_v}}^k\).}

\State {Given \({\bf{\Theta }}_v^{k+1}\), put \(\lambda _{{{\bf{L}}_h}}^{'k} = \xi _{{{\bf{L }}_v}}^k\), solve \eqref{eq21}, retrieve optimal \({\bf{\Theta}}_h^{k+1}\) from \({{{\bf{L }}_h}}\) using Gaussian randomization, and denote the maximized objective value as \(\xi _{{{\bf{L }}_h}}^{k}\).}

\State {Given \({\bf{\Theta }}_v^{k+1}\), \({\bf{\Theta }}_h^{k+1}\), and \({\bf{E }}_h^k\), put \(\lambda _{{{\bf{\Xi }}_v}}^{'k}=\xi _{{{\bf{L }}_h}}^k\), solve \eqref{eq27}, retrieve optimal \({\bf{E}}_v^{k+1}\) from \({{{\bf{\Xi }}_v}}\) using Gaussian randomization, and denote the maximized objective value as \(\xi _{{{\bf{\Xi }}_v}}^k\).}

\State {Given \({\bf{E}}_v^{k+1}\), put \(\lambda _{{{\bf{\Xi }}_h}}^{'k} = \xi _{{{\bf{\Xi }}_v}}^{k}\), solve \eqref{eq29}, retrieve optimal \({\bf{E}}_h^{k+1}\) from \({{{\bf{\Xi }}_h^k}}\) using Gaussian randomization, and denote the maximized objective value as \(\xi _{{{\bf{\Xi }}_h}}^{k+1}\).}

\State {Given \({\bf{\Theta }}_v^{k+1}\), \({\bf{\Theta }}_h^{k+1}\), \({\bf{E }}_h^{k+1}\), and \({\bf{E }}_h^{k+1}\), solve \eqref{eq7} using the SVD and water-filling technique, and denote the optimized \(\bf{W}\), \({\bf{F}}\), and \({{N_s}}\) as \({{\bf{W}}^{k+1}}\), \({{\bf{F}}^{k+1}}\), and \(N_s^{k+1}\), respectively.}

\State {Update \(k = k + 1\)}.

\UNTIL {the objective value of \eqref{eq6} converges or the maximum number of iterations are completed.}
\end{algorithmic}
\end{algorithm}

\section{Proposed Solution to Problem (P1) for Special Cases}

Here, we consider two special scenarios and propose low-complexity solutions to \eqref{eq6}.
\subsection{Low-SNR regime and/or LoS scenario}
In the low-SNR regime and/or LoS scenario, the optimal way to maximize the SE of a MIMO channel is to allocate all the transmit power to the strongest eigenchannel \cite{tse2005fundamentals}. Accordingly, problem \eqref{eq7} for the low-SNR regime can be simplified as
\begin{equation} \tag{P1.24} \label{eq30}
\begin{array}{l}
\mathop {\max }\limits_{{{\bf{\Theta }}_v},{{\bf{\Theta }}_h},{{\bf{E}}_v},{{\bf{E}}_h}} \,\,\,\,{\log _2}\left( {1 + \frac{1}{{{\sigma ^2}}}{{\left| {{\bf{w}}_{n_s^'}^*\left\{ {{{\bf{E}}_h}\left( {{\bf{H}}_{r,v}^H{{\bf{\Theta }}_v}{{\bf{G}}} + {\bf{H}}_{d,v}^H} \right) + {{\bf{E}}_h}\left( {{\bf{H}}_{r,h}^H{{\bf{\Theta }}_h}{{\bf{G}}} + {\bf{H}}_{d,h}^H} \right)} \right\}{\bf{f}}_{n_s^'}^{}} \right|}^2}} \right)\\
{\rm{s}}{\rm{.t}}{\rm{. }}\,\,\,\,\,\,\,\,\,{{\bf{\Theta }}_p} = {\rm{diag}}({e^{ - j{\phi _{p,1}}}},{e^{ - j{\phi _{p,2}}}},{e^{ - j{\phi _{p,3}}}}, \cdots ,{e^{ - j{\phi _{p,N}}}}),\,p \in \left\{ {v,h} \right\}\\
\,\,\,\,\,\,\,\,\,\,\,\,\,\,\,\,{{\bf{E}}_p} = {\rm{diag}}({e^{ - j{\gamma _{p,1}}}},{e^{ - j{\gamma _{p,2}}}},{e^{ - j{\gamma _{p,3}}}}, \cdots ,{e^{ - j{\gamma _{p,{N_r}}}}}),\,p \in \left\{ {v,h} \right\}\\
\,\,\,\,\,\,\,\,\,\,\,\,\,\,\,{\left\| {{\bf{f}}_{n_s^'}^{}} \right\|^2} = {P_t},
\end{array}
\end{equation}
where \({n_s^'}\) denotes the index of the strongest eigenchannel. Note that in \eqref{eq30}, the maximization of the SE is equivalent to the maximization of the signal strength of the \({n_s^'}\)th data stream. Hence, problem \eqref{eq30} can be rewritten as
\begin{equation} \tag{P1.25} \label{eq31}
\begin{array}{l}
\mathop {\max }\limits_{{{\bf{\Theta }}_v},{{\bf{\Theta }}_h},{{\bf{E}}_v},{{\bf{E}}_h}} \,\,\,{\left| {{\bf{w}}_{n_s^'}^*\left\{ {{{\bf{E}}_h}\left( {{\bf{H}}_{r,v}^H{{\bf{\Theta }}_v}{{\bf{G}}} + {\bf{H}}_{d,v}^H} \right) + {{\bf{E}}_h}\left( {{\bf{H}}_{r,h}^H{{\bf{\Theta }}_h}{{\bf{G}}} + {\bf{H}}_{d,h}^H} \right)} \right\}{\bf{f}}_{n_s^'}^{}} \right|^2}\\
{\rm{s}}{\rm{.t}}{\rm{. }}\,\,\,\,\,\,\,\,\,{{\bf{\Theta }}_p} = {\rm{diag}}({e^{ - j{\phi _{p,1}}}},{e^{ - j{\phi _{p,2}}}},{e^{ - j{\phi _{p,3}}}}, \cdots ,{e^{ - j{\phi _{p,N}}}}),\,p \in \left\{ {v,h} \right\},\\
\,\,\,\,\,\,\,\,\,\,\,\,\,\,\,\,{{\bf{E}}_p} = {\rm{diag}}({e^{ - j{\gamma _{p,1}}}},{e^{ - j{\gamma _{p,2}}}},{e^{ - j{\gamma _{p,3}}}}, \cdots ,{e^{ - j{\gamma _{p,{N_r}}}}}),\,p \in \left\{ {v,h} \right\},\\
\,\,\,\,\,\,\,\,\,\,\,\,\,\,\,{\left\| {{\bf{f}}_{n_s^'}^{}} \right\|^2} = {\frac{{{P_t}}}{2}}{\rm{ }}.
\end{array}
\end{equation}

\subsubsection{DP-IRS Phases Optimization for the Low-SNR Regime}
According to \eqref{eq31}, problem \eqref{eq12} for the low-SNR regime is given by
\begin{equation} \tag{P1.26} \label{eq32}
\begin{array}{l}
\mathop {\max }\limits_{{{\bf{l}}_v}} \,\,\,{\left| {{\bf{l}}_v^H{{\bf{{\boldsymbol{\alpha}} }}_{v,n_s^'}}\, + {\eta _{v,n_s^'}}} \right|^2}\\
{\rm{s}}{\rm{.t}}{\rm{. }}\,\,\,\,\,\,\,\,\,\left| {{\bf{l}}_{{v_{\left( n \right)}}}} \right| = 1,\,\,\,\,n = 1, \cdots ,N.
\end{array}
\end{equation}
It can be demonstrated that to solve the problem \eqref{eq32}, we should have \(\arg \left( {{\bf{l}}_v^H{{\boldsymbol{\alpha}} _{v,n_s^'}}} \right) = \arg \left( {{\eta _{v,n_s^'}}} \right)\). Let \(\arg \left( {{\eta _{v,n_s^'}}} \right) = {c_v}\,\). \eqref{eq32} can be reformulated as

\begin{equation} \tag{P1.27} \label{eq33}
\begin{array}{l}
\mathop {\max }\limits_{{{\bf{l}}_v}} \,\,\,{\left| {{\bf{l}}_v^H{{\boldsymbol{\alpha}} _{v,n_s^'}}\,} \right|^2}\\
{\rm{s}}{\rm{.t}}{\rm{. }}\,\,\,\,\,\,\,\,\,\left| {{\bf{l}}_{{v_{\left( n \right)}}}} \right| = 1,\,\,\,\,n = 1, \cdots ,N,\\
\,\,\,\,\,\,\,\,\,\,\,\,\,\,\,\,\arg \left( {{\bf{l}}_v^H{{\boldsymbol{\alpha}} _{v,n_s^'}}} \right) = {c_v}.
\end{array}
\end{equation}
Its not difficult to show that the optimal solution to \eqref{eq33} is \({\bf{l}}_v^H = {e^{j\left( {{c_v} - \arg \left( {{{\bf{{\boldsymbol{\alpha}} }}_{v,n_s^'}}} \right)} \right)}}\) and correspondingly, the \(n\)th phase shift for the vertical part of the DP-IRS is given by
\begin{equation} \label{eq34}
\phi _{{v_{\left( n \right)}}}^* = {c_v} - \arg \left( {{{\bf{{\boldsymbol{\alpha}} }}_{v,n{{_s^'}_{(n)}}}}} \right),\,\,n = 1,2, \cdots ,N.
\end{equation}
Following steps similar to those for \({{{\bf{l}}_v}}\) in \eqref{eq32}, the problem \eqref{eq20} for \({{{\bf{l}}_h}}\) is expressed as
\begin{equation} \tag{P1.28} \label{eq35}
\begin{array}{l}
\mathop {\max }\limits_{{{\bf{l}}_h}} \,\,\,{\left| {{\bf{l}}_h^H{{\boldsymbol{\alpha}} _{h,n_s^'}}\,} \right|^2}\\
{\rm{s}}{\rm{.t}}{\rm{. }}\,\,\,\,\,\,\,\,\,\left| {{\bf{l}}_{{h_{\left( n \right)}}}} \right| = 1,\,\,\,\,n = 1, \cdots ,N,\\
\,\,\,\,\,\,\,\,\,\,\,\,\,\,\,\,\arg \left( {{\bf{l}}_h^H{{\boldsymbol{\alpha}} _{h,n_s^'}}} \right) = {c_h},
\end{array}
\end{equation}
with the optimal solution \({\bf{l}}_h^H = {e^{j\left( {{c_h} - \arg \left( {{{\boldsymbol{\alpha}}_{h,n_s^'}}} \right)} \right)}}\,\), resulting in the following \(n\)th optimal phase shift for the horizontal part of DP-IRS:
\begin{equation} \label{eq36}
\phi _{{h_{\left( n \right)}}}^* = {c_h} - \arg \left( {{{\boldsymbol{\alpha}}_{h,n{{_s^'}_{(n)}}}}} \right),\,n = 1,2, \cdots ,N.
\end{equation}

\subsubsection{Optimal Phase Shifters for Receive DP Antennas}
Corresponding to \eqref{eq31}, problem \eqref{eq24} for \({{{\bf{e}}_v}}\) can be simplified as
\begin{equation} \tag{P1.29} \label{eq37}
\begin{array}{l}
\mathop {\max }\limits_{{{\bf{e}}_v}} \,\,\,\,{\left| {{\bf{e}}_v^{T}{\bf{h}}_{v,n_s^'}^' + \delta _{h,n_s^'}^{}} \right|^2}\\
{\rm{s}}{\rm{.t}}{\rm{. }}\,\,\,\,\,\,\,\,\,{\bf{e}}_v^{T} = \left[ {{e^{ - j{\gamma _{v,1}}}},{e^{ - j{\gamma _{v,2}}}},{e^{ - j{\gamma _{v,3}}}}, \cdots ,{e^{ - j{\gamma _{v,{N_r}}}}}} \right]\\
\,\,\,\,\,\,\,\,\,\,\,\,\,\,\,\,{\bf{h}}_{v,{n_s}}^' = {\rm{diag}}\left( {{\bf{w}}_{n_s^'}^*} \right)\cdot{\bf{h}}_{v,n_s^'}^{}.
\end{array}
\end{equation}
To maximize the objective value of \eqref{eq37} we should have \(\arg \left( {{\bf{e}}_v^{T}{\bf{h}}_{v,n_s^'}^'} \right) = \arg \left( {\delta _{v,n_s^'}^{}} \right)\). Let \(\arg \left( {\delta _{h,n_s^'}^{}} \right) = {u_h}\), the problem \eqref{eq37} is reformulated as
\begin{equation} \tag{P1.30} \label{eq38}
\begin{array}{l}
\mathop {\max }\limits_{{{\bf{e}}_v}} \,\,\,{\left| {{\bf{e}}_v^{T}{\bf{h}}_{v,n_s^'}^'\,} \right|^2}\\
{\rm{s}}{\rm{.t}}{\rm{. }}\,\,\,\,\,\,\,\,\,\left| {{\bf{e}}_{{v_{\left( {{n_r}} \right)}}}} \right| = 1,\,\,\,\,{n_r} = 1, \cdots ,{N_r},\\
\,\,\,\,\,\,\,\,\,\,\,\,\,\,\,\,\arg \left( {{\bf{e}}_v^{T}{\bf{h}}_{v,n_s^'}^'} \right) = {u_h},
\end{array}
\end{equation}
with the optimal solution \({\bf{e}}_v^{T} = {e^{j\left( {{u_h} - \arg \left( {{\bf{h}}_{v,n_s^'}^'} \right)} \right)}}\) and resulting in the following \({n_r}\)th optimal vertical receive phase shift:
\begin{equation} \label{eq39}
{\gamma _{{v_{\left( {n_r} \right)}}}} = {u_h} - \arg \left( {{\bf{h}}_{v,n{{_s^'}_{\left( {n_r} \right)}}}^'} \right),\,\,{n_r} = 1,2, \cdots ,{N_r}.
\end{equation}
Similar to \eqref{eq38}, problem \eqref{eq28} for \({{\bf{e}}_h^{T}}\) is given by
\begin{equation} \tag{P1.31} \label{eq40}
\begin{array}{l}
\mathop {\max }\limits_{{{\bf{e}}_h}} \,\,\,{\left| {{\bf{e}}_h^{T}{\bf{h}}_{h,n_s^'}^'\,} \right|^2}\\
{\rm{s}}{\rm{.t}}{\rm{. }}\,\,\,\,\,\,\,\,\,\left| {{\bf{e}}_{{h_{\left( {{n_r}} \right)}}}} \right| = 1,\,\,\,\,{n_r} = 1, \cdots ,{N_r},\\
\,\,\,\,\,\,\,\,\,\,\,\,\,\,\,\,\arg \left( {{\bf{e}}_h^{T}{\bf{h}}_{h,n_s^'}^'} \right) = {u_v},
\end{array}
\end{equation}
with the optimal solution \({\bf{e}}_h^{T} = {e^{j\left( {{u_v} - \arg \left( {{\bf{h}}_{h,n_s^'}^'} \right)} \right)}}\) and the following \({n_r}\)th optimal horizontal receive phase shift:
\begin{equation} \label{eq41}
{\gamma _{{h_{\left( {n_r} \right)}}}} = {u_v} - \arg \left( {{\bf{h}}_{h,{n_s}_{\left( {n_r} \right)}}^'} \right),\,{n_r} = 1, \cdots ,{N_r},
\end{equation}
where \({u_v} = \arg \left( {\delta _{v,n_s^'}^{}} \right)\).

\subsubsection{Overall Algorithm for the Low-SNR Regime}
The algorithm for the low-SNR regime is summarized in Algorithm \ref{alg:lowSNR}. The convergence of this algorithm is guaranteed by the fact that for each subproblem, the optimal solution is obtained, which ensures that the objective value of \eqref{eq30} is monotonically non-decreasing over iterations. 

\begin{algorithm}[H] \caption{Alternating Optimization Algorithm for the Low-SNR Regime}\label{alg:lowSNR}
\begin{algorithmic}[1]
\State {Set iteration \(k = 0\) and initialize \({{\bf{\Theta }}_v^k} = {{\bf{I}}_{{{\bf{\Theta }}_v}}}\), \({{\bf{\Theta }}_h^k} = {{\bf{I}}_{{{\bf{\Theta }}_h}}}\), \({{\bf{E }}_v^k} = {{\bf{I}}_{{{\bf{E }}_v}}}\), and \({{\bf{E }}_h^k} = {{\bf{I}}_{{{\bf{E }}_h}}}\).}

\State {Solve \eqref{eq7} for strongest eigenchannel and denote the optimized precoding and combining vectors as \({{\bf{f}}_{n_s^{'k}}^{k}}\) and \({{\bf{w}}_{n_s^{'k}}^{*k}}\), respectively.}

\REPEAT
\State {Given \({\bf{E }}_v^k\), \({\bf{E }}_h^k\), and \({\bf{\Theta }}_h^k\), calculate the optimal \({\bf{\Theta }}_v^{k+1}\) using \eqref{eq34}.}

\State {Given \({\bf{\Theta }}_v^{k+1}\), calculate the optimal \({\bf{\Theta }}_h^{k+1}\) using \eqref{eq36}.}

\State {Given \({\bf{\Theta }}_v^{k+1}\), \({\bf{\Theta }}_h^{k+1}\), and \({\bf{E }}_h^k\), calculate the optimal \({\bf{E }}_v^{k+1}\) using \eqref{eq39}.}

\State {Given \({\bf{E }}_v^{k+1}\), calculate the optimal \({\bf{E }}_h^{k+1}\) using \eqref{eq41}.}

\State {Given \({\bf{\Theta }}_v^{k+1}\), \({\bf{\Theta }}_h^{k+1}\), \({\bf{E }}_v^{k+1}\), and \({\bf{E }}_h^{k+1}\), calculate the optimal \({{\bf{f}}_{n_s^{'k+1}}^{k+1}}\), \({{\bf{w}}_{n_s^{'k+1}}^{*k+1}}\), and \({n_s^{'k+1}}\) by solving \eqref{eq7} for strongest eigenchannel.}

\State {Update \(k = k + 1\)}.

\UNTIL {the objective value of \eqref{eq30} converges or the maximum number of iterations are completed.}
\end{algorithmic}
\end{algorithm}

\subsection{High-SNR Regime and Rich Scattering Scenario}

In a high-SNR regime and rich scattering environment, the optimal way to maximize the SE of a MIMO channel is to divide the total transmit power equally to all available eigenchannels \cite{tse2005fundamentals}. Hence, problem \eqref{eq6} can be approximated as \cite{tse2005fundamentals}
\begin{equation} \tag{P1.32} \label{eq42}
\begin{array}{l}
\mathop {\max }\limits_{{{\bf{\Theta }}_v},{{\bf{\Theta }}_h},{{\bf{E}}_v},{{\bf{E}}_h}} \,\,{\sum\limits_{{n_s} = 1}^{{N_s}} {\left| {{\bf{w}}_{{n_s}}^*\left\{ {{{\bf{E}}_h}\left( {{\bf{H}}_{r,v}^H{{\bf{\Theta }}_v}{{\bf{G}}} + {\bf{H}}_{d,v}^H} \right) + {{\bf{E}}_h}\left( {{\bf{H}}_{r,h}^H{{\bf{\Theta }}_h}{{\bf{G}}} + {\bf{H}}_{d,h}^H} \right)} \right\}{\bf{f}}_{{n_s}}^{}} \right|} ^2}\,\,\\
{\rm{s}}{\rm{.t}}{\rm{. }}\,\,\,\,\,\,\,\,\,{{\bf{\Theta }}_p} = {\rm{diag}}({e^{ - j{\phi _{p,1}}}},{e^{ - j{\phi _{p,2}}}},{e^{ - j{\phi _{p,3}}}}, \cdots ,{e^{ - j{\phi _{p,N}}}}),\,p \in \left\{ {v,h} \right\},\\
\,\,\,\,\,\,\,\,\,\,\,\,\,\,\,\,{{\bf{E}}_p} = {\rm{diag}}({e^{ - j{\gamma _{p,1}}}},{e^{ - j{\gamma _{p,2}}}},{e^{ - j{\gamma _{p,3}}}}, \cdots ,{e^{ - j{\gamma _{p,{N_r}}}}}),\,p \in \left\{ {v,h} \right\},\\
\,\,\,\,\,\,\,\,\,\,\,\,\,\,\,{\left\| {{\bf{f}}_{{n_s}}^{}} \right\|^2} = \frac{{{P_t}}}{{{N_s}}},\,{n_s} = 1,2, \cdots ,{2N_s}.
\end{array}
\end{equation}
Following the same method as in Section II, the SDR formulations for optimizing \({{\bf{\Theta }}_v}\), \({{\bf{\Theta }}_h}\), \({{\bf{E }}_v}\), and \({{\bf{E }}_h}\) are given by
\begin{equation} \tag{P1.33} \label{eq43} 
\begin{array}{l}
\mathop {\max }\limits_{{\bf{L}}_v^{}} \,\,\,\,{\xi _{{\bf{L}}_v^{}}}\\
{\rm{s}}{\rm{.t}}{\rm{. }}\,\,\,\,\,\,\,\,\,\,{{\bf{x}}_{{\bf{L}}_v^{}}} = \left[ {\left( {{\rm{trace}}\left( {{\bf{O}}_{v,1}^{}{\bf{L}}_v^{}} \right) + {{\left| {{\eta _{v,1}}} \right|}^2}} \right),\, \cdots ,\left( {{\rm{trace}}\left( {{\bf{O}}_{v,{n_s}}^{}{\bf{L}}_v^{}} \right) + {{\left| {{\eta _{v,{n_s}}}} \right|}^2}} \right)} \right]\\
\,\,\,\,\,\,\,\,\,\,\,\,\,\,\,\,\,{{\bf{L}}_{{v_{\left( {n,n} \right)}}}} = 1,\,\,\,\,n = 1, \cdots ,N + 1,\,\\
\,\,\,\,\,\,\,\,\,\,\,\,\,\,\,\,\,{{\bf{L}}_v}\underline  \succ  0,\\
\,\,\,\,\,\,\,\,\,\,\,\,\,\,\,\,\,{\rm{sum}}\left( {{\bf{x}}_{{\bf{L}}_v^{}}} \right) \ge \lambda _{{\bf{L}}_v^{}}^' + {\xi _{{\bf{L}}_v^{}}},
\end{array}
\end{equation}
\begin{equation} \tag{P1.34} \label{eq44}
\begin{array}{l}
\mathop {\max }\limits_{{\bf{L}}_h^{}} \,\,\,\,{\xi _{{\bf{L}}_h^{}}}\\
{\rm{s}}{\rm{.t}}{\rm{. }}\,\,\,\,\,\,\,\,\,\,{{\bf{x}}_{{\bf{L}}_h^{}}} = \left[ {\left( {{\rm{trace}}\left( {{\bf{O}}_{h,1}^{}{\bf{L}}_h^{}} \right) + {{\left| {{\eta _{h,1}}} \right|}^2}} \right),\, \cdots ,\left( {{\rm{trace}}\left( {{\bf{O}}_{h,{n_s}}^{}{\bf{L}}_h^{}} \right) + {{\left| {{\eta _{h,{n_s}}}} \right|}^2}} \right)} \right]\\
\,\,\,\,\,\,\,\,\,\,\,\,\,\,\,\,\,\,{{\bf{L}}_{{h_{\left( {n,n} \right)}}}} = 1,\,\,\,\,n = 1, \cdots ,N + 1,\,\\
\,\,\,\,\,\,\,\,\,\,\,\,\,\,\,\,\,\,{{\bf{L}}_h}\underline  \succ  0,\\
\,\,\,\,\,\,\,\,\,\,\,\,\,\,\,\,\,\,{\rm{sum}}\left( {{\bf{x}}_{{\bf{L}}_h^{}}} \right) \ge \lambda _{{{\bf{L}}_h}}^' + {\xi _{{\bf{L}}_h^{}}},
\end{array}
\end{equation}
\begin{equation} \tag{P1.35} \label{eq45}
\begin{array}{l}
\mathop {\max }\limits_{{{\bf{\Xi}} _v}} \,\,\,\,{\xi _{{{\bf{\Xi}} _v}}}\\
{\rm{s}}{\rm{.t}}{\rm{. }}\,\,\,\,\,\,\,\,\,\,{{\bf{x}}_{{\bf{\Xi}}_v^{}}} = \left[ {\left( {{\rm{trace}}\left( {{\bf{S}}_1^v{{\bf{\Xi}} _v}} \right) + {{\left| {{\gamma _h}} \right|}^2}} \right),\, \cdots ,\,\left( {{\rm{trace}}\left( {{\bf{S}}_{{n_s}}^v{{\bf{\Xi}} _v}} \right) + {{\left| {{\gamma _h}} \right|}^2}} \right)} \right]\\
\,\,\,\,\,\,\,\,\,\,\,\,\,\,\,\,\,{{\bf{\Xi }}_{{v_{\left( {{n_r},{n_r}} \right)}}}} = 1,\,\,n = 1, \cdots ,{N_r} + 1,\\
\,\,\,\,\,\,\,\,\,\,\,\,\,\,\,\,\,{{\bf{\Xi}} _v}\underline  \succ  0,\\
\,\,\,\,\,\,\,\,\,\,\,\,\,\,\,\,{\rm{sum}}\left( {{\bf{x}}_{{\bf{\Xi}}_v^{}}} \right) \ge \lambda _{{{\bf{\Xi}} _v}}^' + {\xi _{{{\bf{\Xi}} _v}}},
\end{array}
\end{equation}
and
\begin{equation} \tag{P1.36} \label{eq46} 
\begin{array}{l}
\mathop {\max }\limits_{{{\bf{\Xi}} _h}} \,\,\,\,{\xi _{{{\bf{\Xi}} _h}}}\\
{\rm{s}}{\rm{.t}}{\rm{. }}\,\,\,\,\,\,\,\,\,\,{{\bf{x}}_{{\bf{\Xi}}_h^{}}} = \left[ {\left( {{\rm{trace}}\left( {{\bf{S}}_1^h{{\bf{\Xi}} _h}} \right) + {{\left| {{\gamma _v}} \right|}^2}} \right),\, \cdots ,\,\left( {{\rm{trace}}\left( {{\bf{S}}_{{n_s}}^h{{\bf{\Xi}} _h}} \right) + {{\left| {{\gamma _v}} \right|}^2}} \right)} \right]\\
\,\,\,\,\,\,\,\,\,\,\,\,\,\,\,\,\,{{\bf{\Xi }}_{{h_{\left( {{n_r},{n_r}} \right)}}}} = 1,\,\,n = 1, \cdots ,{N_r} + 1,\\
\,\,\,\,\,\,\,\,\,\,\,\,\,\,\,\,\,{{\bf{\Xi}} _h}\underline  \succ  0\\
\,\,\,\,\,\,\,\,\,\,\,\,\,\,\,\,\,{\rm{sum}}\left( {{\bf{x}}_{{\bf{\Xi}}_h^{}}} \right) \ge \lambda _{{{\bf{\Xi}} _h}}^' + {\xi _{{{\bf{\Xi}} _h}}},
\end{array}
\end{equation}
respectively. The AO algorithm based on these formulations requires a lower computational cost than that of the AO in Section II for the following reasons: 1) There is no need to adopt the water-filling method in each AO iteration. 2) Solving problems based on \({\rm{sum}}\left( {\bf{x}} \right)\) is computationally less expensive than those based on \({\rm{GM}}\left( {\bf{x}} \right)\) \cite{cvx}. Furthermore, in the high-SNR regime, the expressions for \({\bf{x}}\) in \eqref{eq43}, \eqref{eq44}, \eqref{eq45}, and \eqref{eq46} are simpler than those in Section II. The overall algorithm is summarized in Algorithm \ref{alg:highSNR}.

If the environment is not favorable for Algorithm \ref{alg:highSNR}, the water-filling power allocation method in Algorithm \ref{alg:highSNR} can result in better performance than Algorithm \ref{alg:highSNR} with equal power distribution. The performance analysis of Algorithm \ref{alg:highSNR} with optimal power allocation is also presented in Section \ref{Sect:SimulationResults}. 
\begin{algorithm}[H] \caption{Alternating Optimization Algorithm for High SNR regime}\label{alg:highSNR}
\begin{algorithmic}[1]
\State {Set iteration \(k  = 0\) and initialize \({{\bf{\Theta }}_v^k} = {{\bf{I}}_{{{\bf{\Theta }}_v}}}\),  \({{\bf{\Theta }}_h^k} = {{\bf{I}}_{{{\bf{\Theta }}_h}}}\), \({{\bf{E }}_v^k} = {{\bf{I}}_{{{\bf{E }}_v}}}\), \({{\bf{E }}_h^k} = {{\bf{I}}_{{{\bf{E }}_h}}}\), and \(\xi _{{{\bf{\Xi }}_h}}^{k}=0\).}

\State {Solve \eqref{eq7} using the SVD, divide the total transmit power equally among the available eigenchannels, and denote the optimized \(\bf{W}\), \({\bf{F}}\), and \({{N_s}}\) as \({{\bf{W}}^k}\), \({{\bf{F}}^k}\), and \(N_s^k\), respectively.}

\REPEAT

\State {Given \({\bf{E }}_v^{k}\), \({\bf{E }}_h^{k}\), and \({\bf{\Theta }}_h^k\), put \(\lambda _{{{\bf{L}}_v}}^{'k} = \xi _{{{\bf{\Xi }}_h}}^k\), solve \eqref{eq43}, retrieve optimal \({\bf{\Theta}}_v^{k+1}\) from \({{{\bf{L }}_v}}\) using Gaussian randomization, and denote the maximized objective value as \(\xi _{{{\bf{L }}_v}}^k\).}

\State {Given \({\bf{\Theta }}_v^{k+1}\), put \(\lambda _{{{\bf{L}}_h}}^{'k} = \xi _{{{\bf{L }}_v}}^k\), solve \eqref{eq44}, retrieve optimal \({\bf{\Theta}}_h^{k+1}\) from \({{{\bf{L }}_h}}\) using Gaussian randomization, and denote the maximized objective value as \(\xi _{{{\bf{L }}_h}}^{k}\).}

\State {Given \({\bf{\Theta }}_v^{k+1}\), \({\bf{\Theta }}_h^{k+1}\), and \({\bf{E }}_h^k\), put \(\lambda _{{{\bf{\Xi }}_v}}^{'k}=\xi _{{{\bf{L }}_h}}^k\), solve \eqref{eq45}, retrieve optimal \({\bf{E}}_v^{k+1}\) from \({{{\bf{\Xi }}_v}}\) using Gaussian randomization, and denote the maximized objective value as \(\xi _{{{\bf{\Xi }}_v}}^k\).}

\State {Given \({\bf{E}}_v^{k+1}\), put \(\lambda _{{{\bf{\Xi }}_h}}^{'k} = \xi _{{{\bf{\Xi }}_v}}^{k}\), solve \eqref{eq46}, retrieve optimal \({\bf{E}}_h^{k+1}\) from \({{{\bf{\Xi }}_h^k}}\) using Gaussian randomization, and denote the maximized objective value as \(\xi _{{{\bf{\Xi }}_h}}^{k+1}\).}

\State {Given \({\bf{\Theta }}_v^{k+1}\), \({\bf{\Theta }}_h^{k+1}\), \({\bf{E }}_h^{k+1}\), and \({\bf{E }}_h^{k+1}\), solve \eqref{eq7} using the SVD, divide the total transmit power equally among the available eigenchannels, and denote the optimized \(\bf{W}\), \({\bf{F}}\), and \({{N_s}}\) as \({{\bf{W}}^{k+1}}\), \({{\bf{F}}^{k+1}}\), and \(N_s^{k+1}\), respectively.}

\State {Update \(k = k + 1\)}.

\UNTIL {the objective value of \eqref{eq42} converges or the maximum number of iterations are completed.}
\end{algorithmic}
\end{algorithm}

\section{Simulation Results}

\subsection{Simulation Setup}

\begin{figure}[h]
\centering
\includegraphics[width=0.48\textwidth]{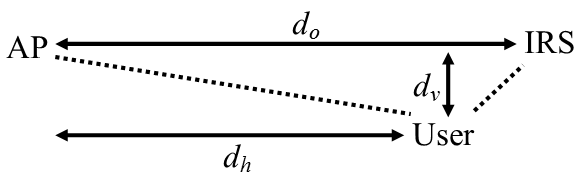}
\caption{Simulation setup.}
\label{fig:SystemSetup}
\end{figure}
In this section, we provide the numerical analysis results to determine the performance of the proposed algorithms. The simulation setup, which is widely used for research, is illustrated in Fig. \ref{fig:SystemSetup} \cite{wu2019intelligent,wu2019beamforming,wu2018intelligent}. Owing to the moving capability of the user, we consider Rayleigh fading for the AP-user and IRS-user links, whereas the AP-IRS link is modeled by Rician fading \cite{viswanathan2020wireless}. Specifically, the AP-IRS channel is given by
\begin{equation}\label{eq47}
{{\bf{G}}_p} = \sqrt {\frac{\chi }{{1 + \chi }}} {\bf{G}}_p^{LoS} + \sqrt {\frac{1}{{1 + \chi }}} {\bf{G}}_p^{NLoS},\,\,p \in \left\{ {v,h} \right\},
\end{equation}
where \({\bf{G}}_p^{LoS}\) and \({\bf{G}}_p^{NLoS}\) represent LoS and non-LoS (i.e., Rayleigh) components of \({{\bf{G}}_p}\), respectively, and \(\chi \) denotes the Rician factor. The variation in the value of \(\chi \) determines the dominant type of AP-IRS link, i.e., LoS or Rayleigh. The AP-IRS distance (\({d_o}\)) and IRS-user vertical distance (\({d_v}\)) are fixed at 40 m and 2 m, respectively. The AP-user horizontal distance, denoted by \({d_h}\), is varied to observe the change in performance. After setting \({d_h}\), \({d_o}\), and \({d_v}\), we can calculate the direct distances between the AP, IRS, and the user. To determine the path loss for each link, we use
\begin{equation} \label{eq48}
PL({\rm{dB}}) = {C_o} + 10a\log \left( {\frac{d}{{{D_o}}}} \right),
\end{equation}
where \(a\) is the path loss exponent, \({C_o}\) is the path loss (\({\rm{dB}}\)) at a reference distance of \({D_o}\), and \(d\) is the distance between links. According to \cite{wu2019intelligent}, we set \({D_o} = 1\, {\rm{m}}\), \({C_o} = 30\, {\rm{dB}}\), \({P_t} = 40 \,{\rm{dBm}}\), \({\sigma^2} = -94\, {\rm{dBm}}\), and \(N_t = N_r = 4\) unless otherwise stated. Furthermore, we assume \({a} = \) 2.2, 3.5, and 2.5 for the AP-IRS, AP-User, and IRS-User links, respectively, \cite{wu2019intelligent}. All simulation results are averaged over 100 channel realizations and performed using MATLAB R2022a on an AMD Ryzen R7-5800H CPU @3.20GHz with 16 GB of RAM.

\subsection{Simulation Results}
\label{Sect:SimulationResults}
\begin{figure}[h]
\centering
\includegraphics[width=0.5\textwidth]{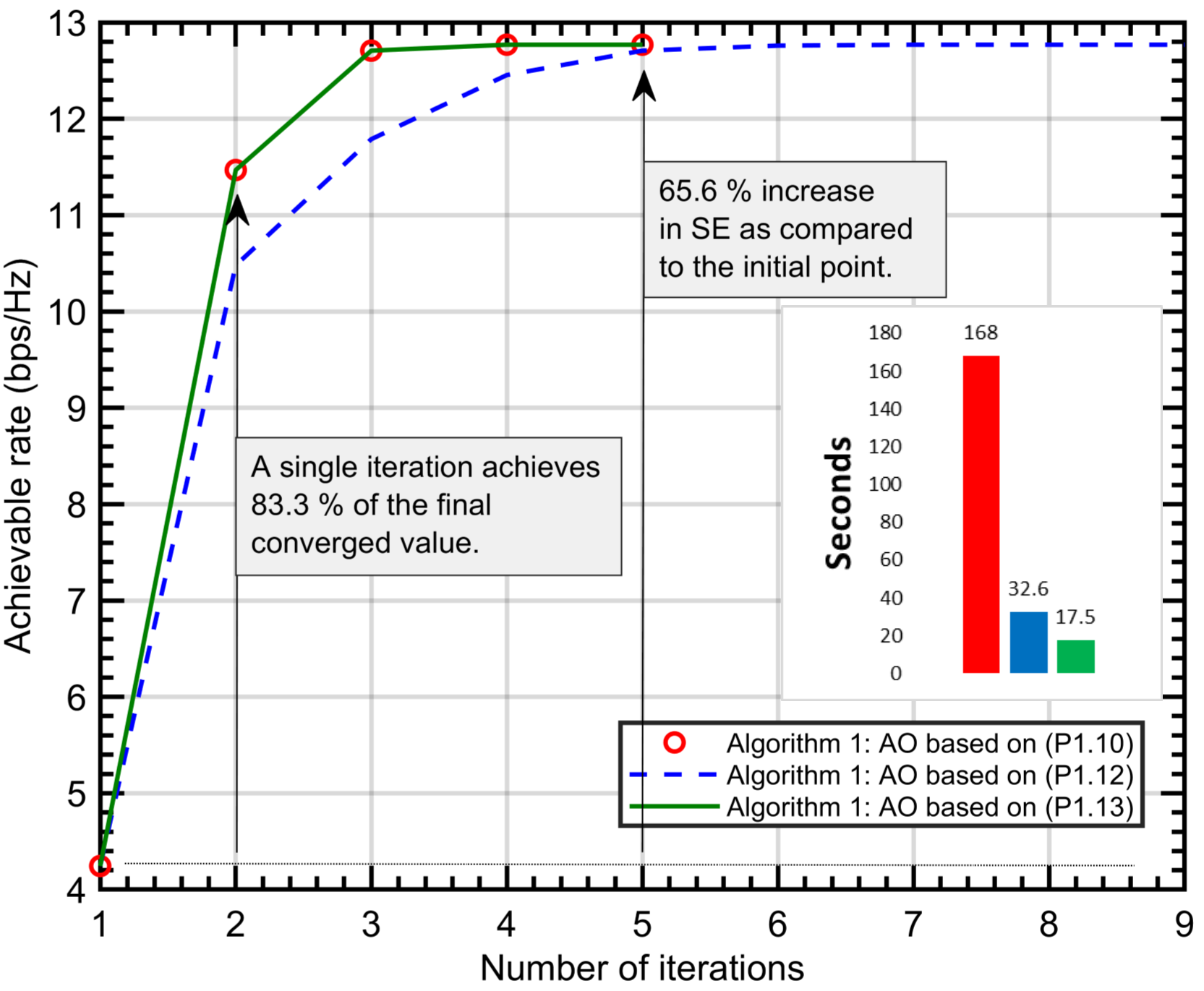}
\caption{Convergence behavior of Algorithm 1 with \(N = 50\). }
\label{fig:1Convergence}
\end{figure}

First, we set \(N = 50\), \({d_h} = 38\) meters, and \(\chi = -20\,\rm{dB}\), and plot the convergence behavior and show the time complexity of Algorithm 1 in Fig. \ref{fig:1Convergence}. Specifically, in Fig. \ref{fig:1Convergence}, Algorithm 1 is simulated based on the following three formulations:
\begin{enumerate}
    \item AO based on \eqref{eq17a} formulation: Algorithm \ref{alg:sdr} solves \eqref{eq17a} and it can be observed that it requires a significantly higher computational cost than the other two formulations because of \({\rm{logsum}}\left( {\bf{x}} \right)\) expressions \cite{cvxProblem}.  
    
    \item AO based on \eqref{eq18} formulation: Algorithm \ref{alg:sdr} solves \eqref{eq18} and it can be observed that it takes more iterations than Case 2 because there is no objective function.  
    
    \item AO based on \eqref{eq19} formulation: Algorithm 1 solves \eqref{eq19}, where a slack variable and \({\rm{GM}}\left( {\bf{x}} \right)\) serve as an alternative to the objective function and \({\rm{logsum}}\left( {\bf{x}} \right)\), respectively. It can be observed in Fig. \ref{fig:1Convergence} that Algorithm \ref{alg:sdr}, based on this formulation, requires the lowest computational cost. 
    
\end{enumerate}
Furthermore, in Fig. \ref{fig:1Convergence}, it can be seen that the Algorithm \ref{alg:sdr} provides 65.6 \% increase in SE compared to that of the initial point and the first iteration achieves 83.3 \% of the converged value.

In Table \ref{tabflops}, we present the computational complexity of Algorithms 1, 2, and 3 in terms of the average CPU running time. It can be inferred that Algorithms 2 and 3 achieve significant reductions in computational complexity compared with Algorithm 1. Specifically, for \(N = 50\) and \(16 \times 16\) MIMO, Algorithms 1, 2, and 3 take 19.57 s, 10.02 ms, and 4.19 s, respectively. Furthermore, in Table \ref{tabflops}, it can be noted that because of the closed-form solutions, Algorithm 2 is significantly faster; in other words, it requires only several milliseconds for convergence.
\begin{table}[h] \centering
\caption{Average CPU running time for Algorithms 1, 2, and 3.}\label{tabflops}
\begin{tabular}{|c|ccc|ccc|}
\hline
\multirow{2}{*}{\textbf{Algorithm}} & \multicolumn{1}{c|}{\textbf{\(N = 10\)}} & \multicolumn{1}{c|}{\textbf{\(N = 30\)}} & \textbf{\(N = 50\)} & \multicolumn{1}{c|}{\textbf{\(N = 10\)}} & \multicolumn{1}{c|}{\textbf{\(N = 30\)}} & \textbf{\(N = 50\)} \\ \cline{2-7} 
                                    & \multicolumn{3}{c|}{\(4 \times 4\,{\rm{MIMO}}\)}                                                               & \multicolumn{3}{c|}{\(16 \times 16\,{\rm{MIMO}}\)}                                                             \\ \hline
Algorithm 1                         & \multicolumn{1}{c|}{1.83 s}             & \multicolumn{1}{c|}{3.17 s}             & 17.51 s             & \multicolumn{1}{c|}{2.96 s}             & \multicolumn{1}{c|}{5.02 s}             & 19.57 s             \\ \hline
Algorithm 2                         & \multicolumn{1}{c|}{5.82 ms}               & \multicolumn{1}{c|}{7.11 ms}               & 9.40 ms               & \multicolumn{1}{c|}{7.13 ms}               & \multicolumn{1}{c|}{8.27 ms}               & 10.02 ms               \\ \hline
Algorithm 3                         & \multicolumn{1}{c|}{1.12 s}              & \multicolumn{1}{c|}{2.05 s}             & 6.61 s            & \multicolumn{1}{c|}{1.61 s}              & \multicolumn{1}{c|}{2.22 s}             & 4.19 s             \\ \hline
\end{tabular}
\end{table}

\begin{figure}[h]
\centering
\includegraphics[width=0.5\textwidth]{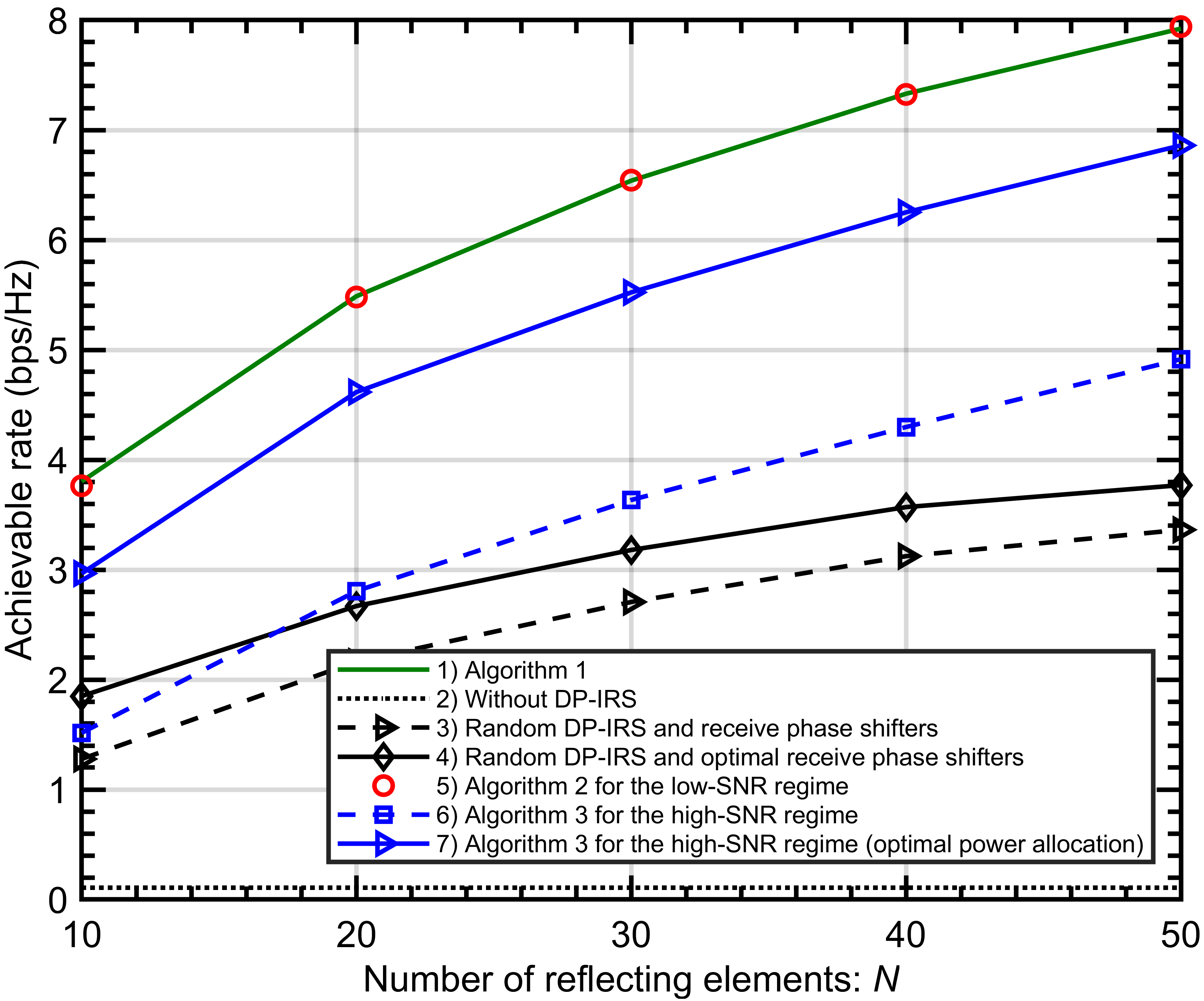}
\caption{Achievable rate versus the number of reflecting elements for \(\chi = 20\,\rm{dB}\).}
\label{fig:2SE vs N LoS}
\end{figure}

\begin{figure}[h]
\centering
\includegraphics[width=0.5\textwidth]{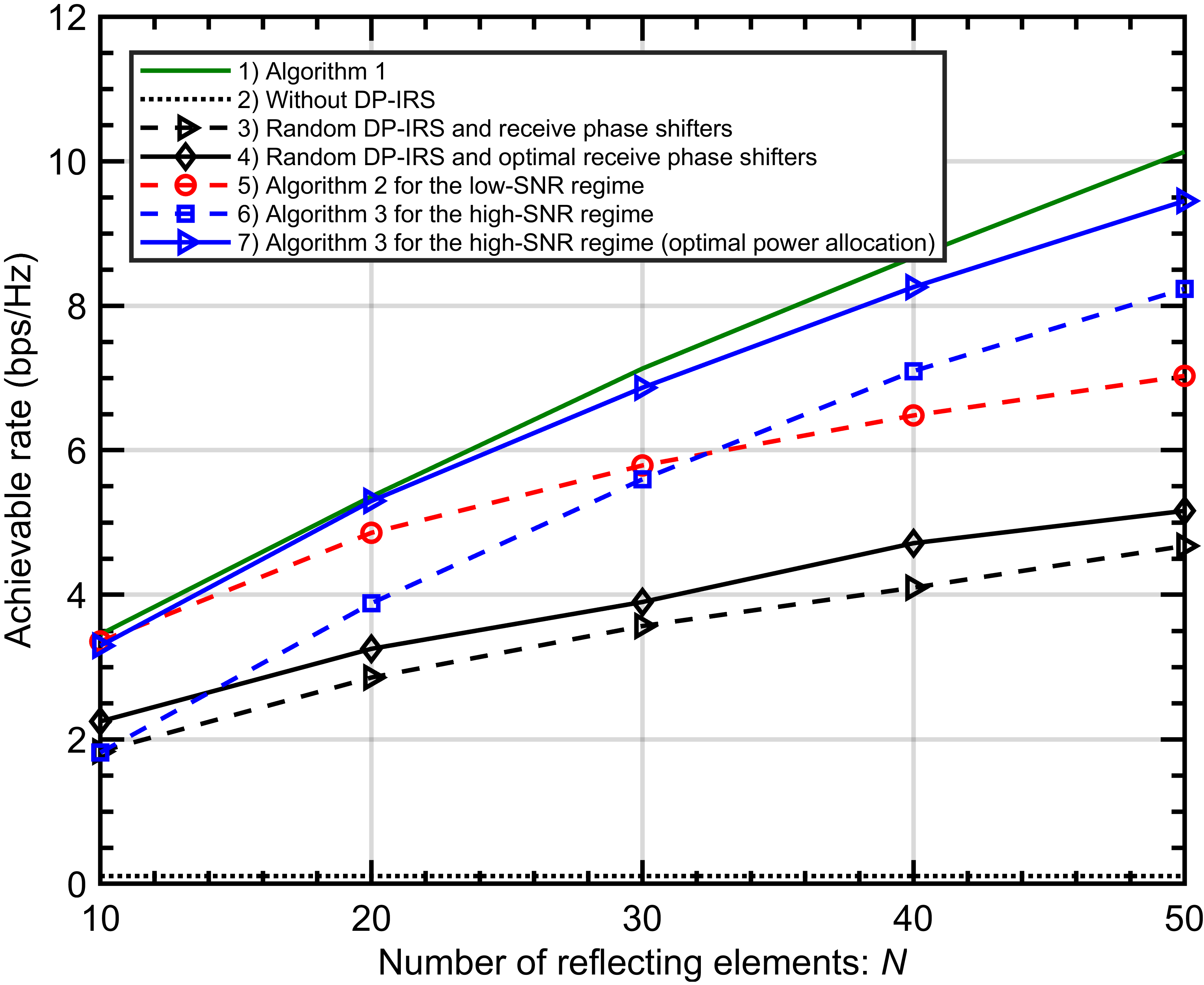}
\caption{Achievable rate versus the number of reflecting elements for \(\chi = - 20\,\rm{dB}\).}
\label{fig:3SE vs N Rayeligh}
\end{figure}

Next, in Figs. \ref{fig:2SE vs N LoS} and \ref{fig:3SE vs N Rayeligh}, we plot the SE performance versus the number of reflecting elements for \(\chi = 20\,\rm{dB}\) and \(-20\,\rm{dB}\), respectively. The comparison schemes are as follows.

\begin{enumerate}
    \item Algorithm 1: In Figs. \ref{fig:2SE vs N LoS} and \ref{fig:3SE vs N Rayeligh}, it can be observed that Algorithm \ref{alg:sdr}, i.e., SDR based AO, achieves the best performance in both scenarios, i.e., Rayleigh and LoS.
    
    \item Without DP-IRS: There is no IRS and the precoder \({\bf{F}}\) and combiner \({\bf{W}}\) are optimized based on SVD and the water-filling technique. It is observed that without IRS, the user achieves the worst SE performance.
    
    \item Random DP-IRS and receive phase shifters: We consider random phases for both DP-IRS and receive phase shifters. In Figs. \ref{fig:2SE vs N LoS} and \ref{fig:3SE vs N Rayeligh}, it can be observed that the random operations result in the significantly lower SE performance as compared to that of Algorithm 1.
    
    \item Random DP-IRS and optimized receive phase shifters: To demonstrate the impact of optimal receive phase shifters on the SE performance, in Algorithm 1, we consider random DP-IRS phases and optimal receive phase shifts. It can be observed that even if the DP-IRS operations are random, optimizing only receive phase shifters improves SE performance.   
    
    \item Algorithm 2: We solve the problem \eqref{eq6} using Algorithm 2, which is designed for the low-SNR regime and/or LoS scenarios. It can be observed that for \(\chi = 20\,\rm{dB}\) in Fig. \ref{fig:2SE vs N LoS}, Algorithm 2 achieves almost the same performance as Algorithm 1 with significantly lower computational cost, as presented in Table \ref{tabflops}. Furthermore, in Fig. \ref{fig:3SE vs N Rayeligh}, for \(\chi = -20\,\rm{dB}\), i.e., Rayleigh fading, Algorithm 2 still performs very close to Algorithm \ref{alg:lowSNR} when \(N\) is small. However, as \(N\) increases, the performance difference between Algorithms 1 and 2 becomes larger because large values of \(N\) result in high SNRs.      
    
    \item Algorithm 3: We consider Algorithm \ref{alg:highSNR}, which is designed for high-SNR and rich-scattering environments. This algorithm exhibits optimal performance only when the powers of all eigenchannels are equally strong \cite{tse2005fundamentals}. In Fig. \ref{fig:3SE vs N Rayeligh}, it can be observed that as \(N\) increases, i.e., varying from the low-SNR regime to the high-SNR regime, the SE performance of Algorithm \ref{alg:highSNR} gets better than that of Algorithm 2. However, in Fig. \ref{fig:2SE vs N LoS}, where the dominant effective channel is LoS, Algorithm 3 results in poor performance.
    
    \item Algorithm 3 with optimal power distribution: We consider algorithm 3 with optimal power distribution. It can be observed in Figs. \ref{fig:2SE vs N LoS} and \ref{fig:3SE vs N Rayeligh} that Algorithm 3 with optimal power distribution achieves a significantly better performance. Specifically, in Fig. \ref{fig:3SE vs N Rayeligh}, where \(\chi = -20\,\rm{dB}\), it performs close to Algorithm \ref{alg:sdr}. 
    
\end{enumerate}

\begin{figure}[h]
\centering
\includegraphics[width=0.5\textwidth]{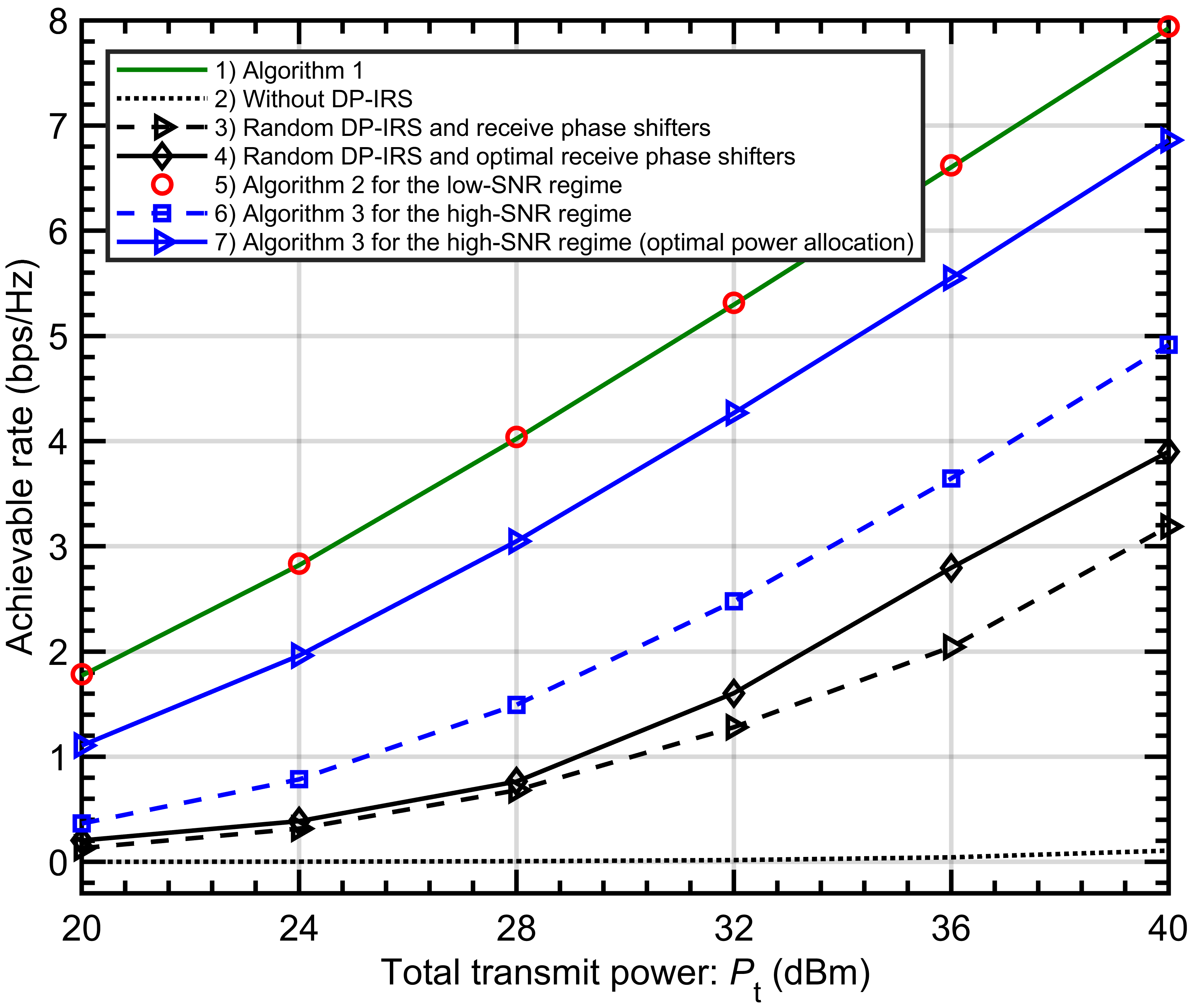}
\caption{Achievable rate versus the total transmit power for \(\chi = 20\,\rm{dB}\).}
\label{fig:4SE vs Power LoS}
\end{figure}

\begin{figure}[h]
\centering
\includegraphics[width=0.5\textwidth]{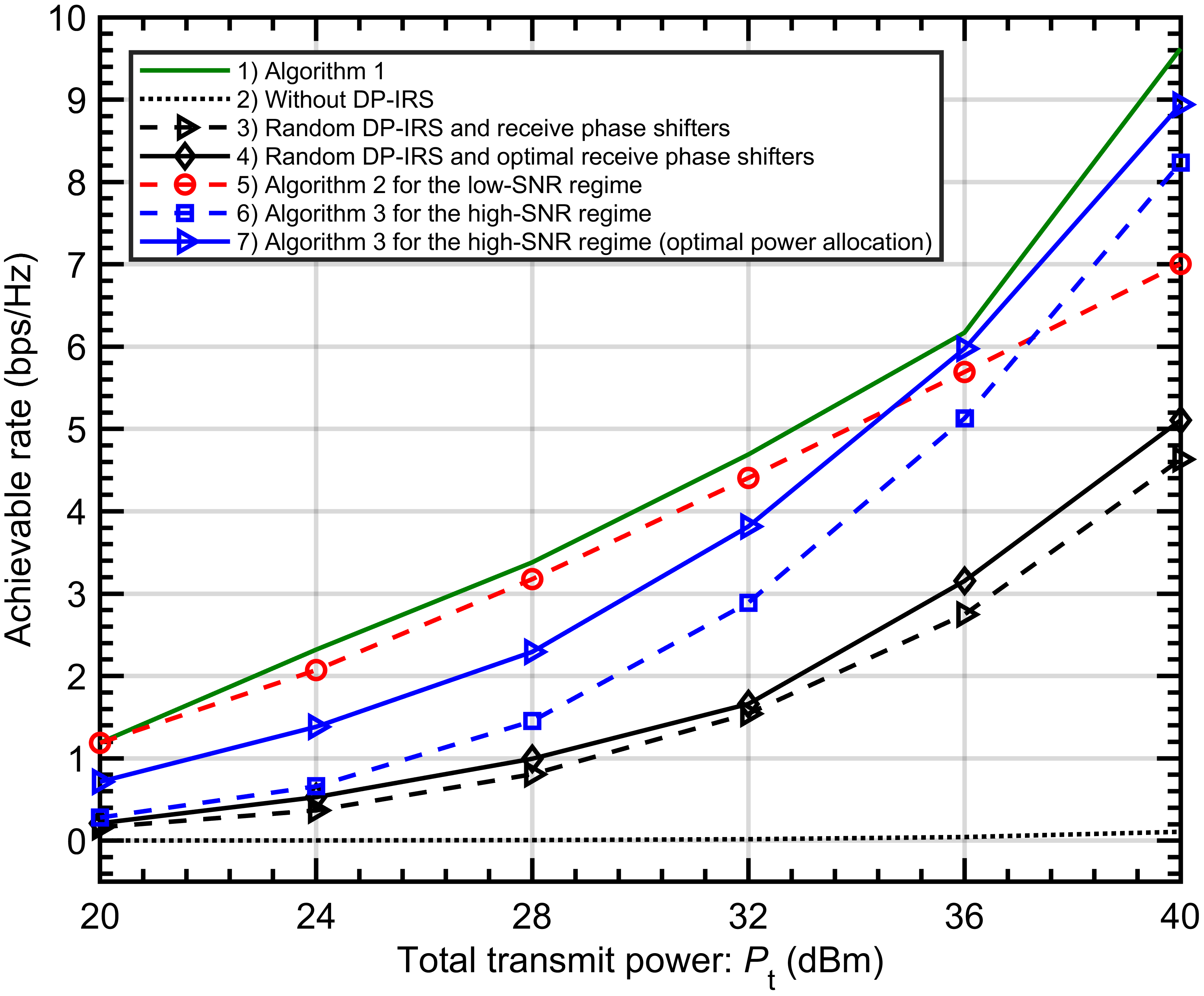}
\caption{Achievable rate versus the total transmit power for \(\chi = - 20\,\rm{dB}\).}
\label{fig:5 SE vs power Rayleigh}
\end{figure}

Next, in Figs. \ref{fig:4SE vs Power LoS} and \ref{fig:5 SE vs power Rayleigh}, we plot the SE performance versus the total transmit power for \(\chi = 20\,\rm{dB}\) and \(-20\,\rm{dB}\), respectively. Again, Algorithm 1 achieves the best performance in both the figures. In general, similar relative performance behaviors are observed, as illustrated in Figs. \ref{fig:2SE vs N LoS} and \ref{fig:3SE vs N Rayeligh}.

\begin{figure}[h]
\centering
\includegraphics[width=0.5\textwidth]{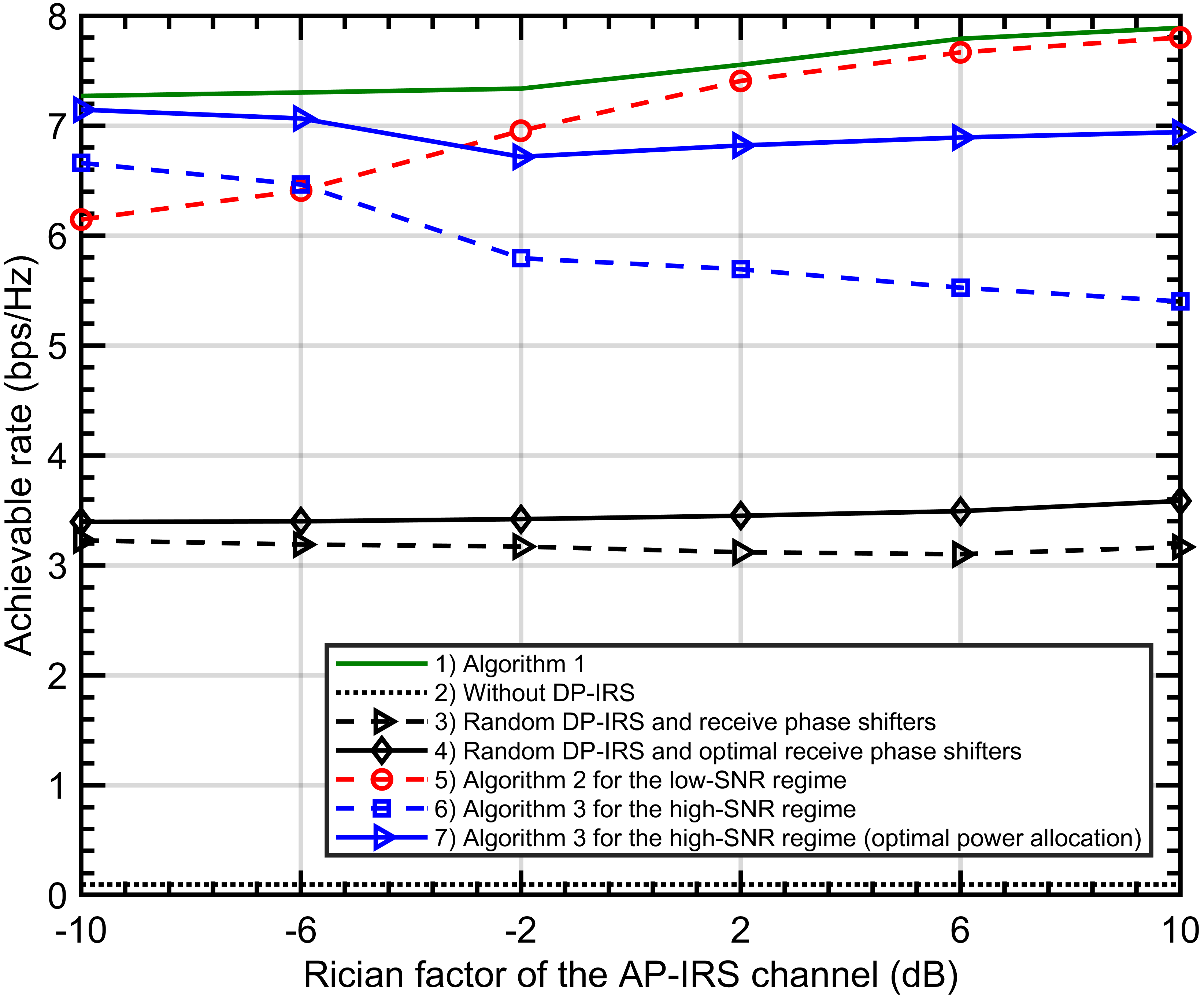}
\caption{Achievable rate versus the Rician factor of AP-IRS link.}
\label{fig:6 SE vs Rician}
\end{figure}

In Fig. \ref{fig:6 SE vs Rician}, we demonstrate the SE performance versus the Rician factor \(\chi\). Specifically, for \(N = 50\) and \({d_h} = 38\) m, we vary \(\chi\) from \(-10\) dB to \(10\) dB, i.e., from Rayleigh to LoS. In Fig. \ref{fig:6 SE vs Rician}, it can be observed that as the channel changes from Rayleigh to LoS, the SE performance of Algorithm 2 improves. The opposite behavior can be observed in Algorithm 3. Overall, Algorithm 1 maintains the best SE performance for all values of \(\chi\). 

\begin{figure}[h]
\centering
\includegraphics[width=0.5\textwidth]{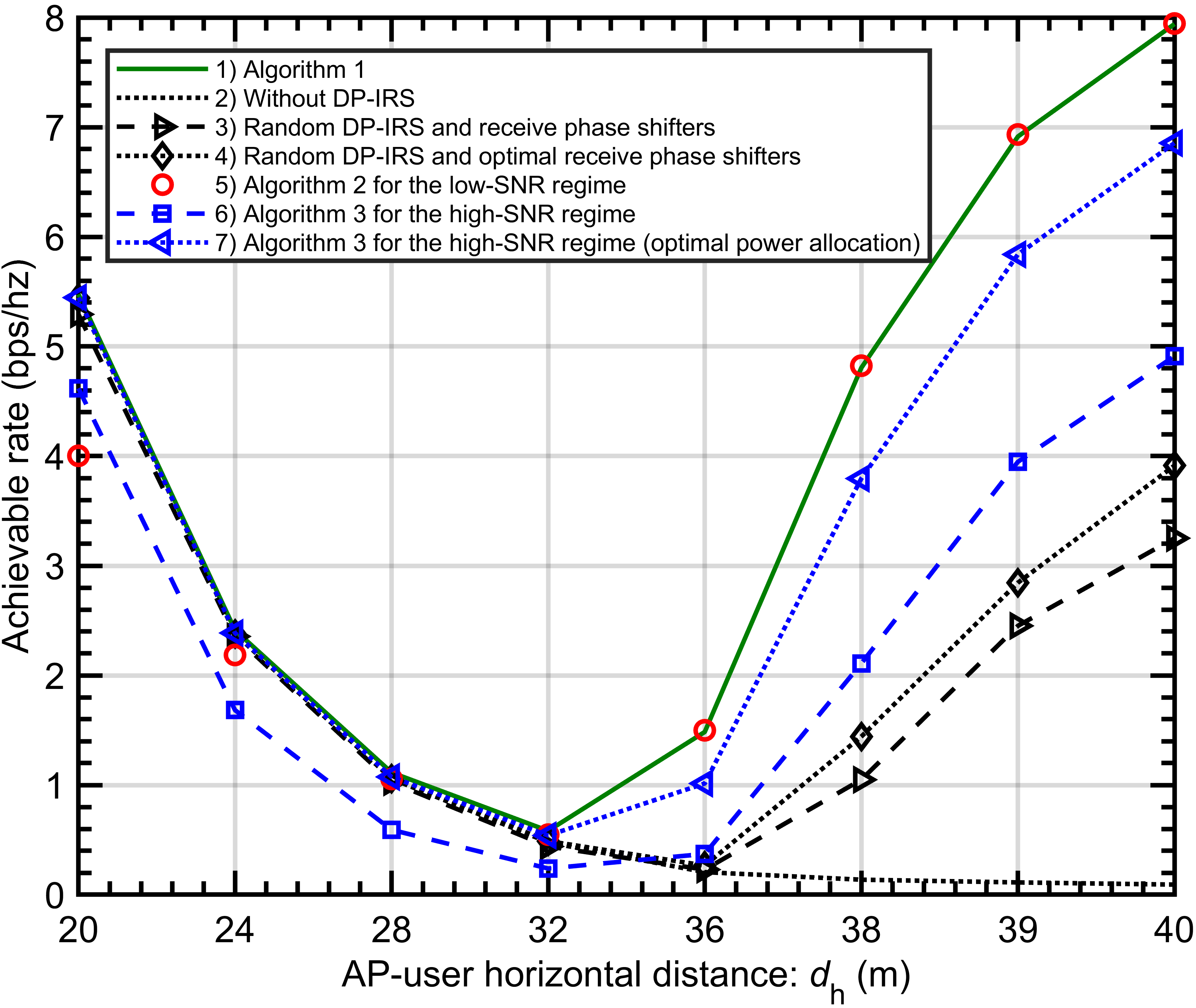}
\caption{Achievable rate versus the AP-user horizontal distance for \(\chi = 20\,\rm{dB}\).}
\label{fig:7 SE vs distance Rayleigh}
\end{figure}

\begin{figure}[h]
\centering
\includegraphics[width=0.5\textwidth]{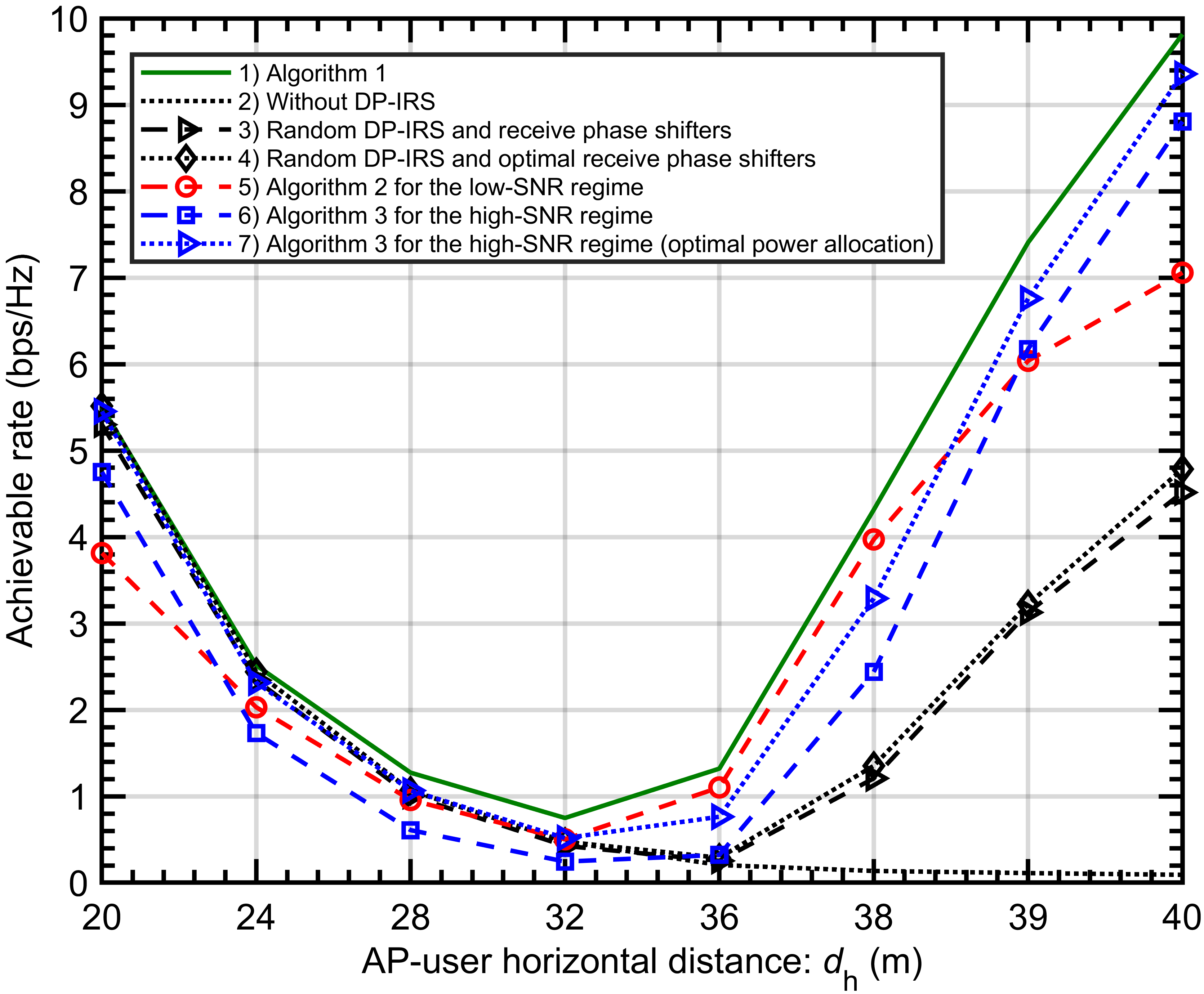}
\caption{Achievable rate versus the AP-user horizontal distance for \(\chi = - 20\,\rm{dB}\).}
\label{fig8:SE vs distance Rayleigh}
\end{figure}

Next, in Figs. \ref{fig:7 SE vs distance Rayleigh} and \ref{fig8:SE vs distance Rayleigh}, we move the user from the AP to the IRS and plot the resulting SE performance for \(\chi = 20\,\rm{dB}\) and \(-20\,\rm{dB}\), respectively. Specifically, for \(N = 50\), we alter \({d_h}\) from 20 m to 40 m. In Figs. \ref{fig:7 SE vs distance Rayleigh} and \ref{fig8:SE vs distance Rayleigh}, when the user is near the AP and far from the DP-IRS, i.e., \({d_h} < 32\)m, the SE performance is mainly determined by the AP-user direct link. Hence, it can be observed that schemes employing the optimal power allocation method, i.e., Schemes 1, 2, 3, 4, and 7, achieve similar SE performance. In contrast, the SE performance is lower in the other schemes, i.e., Schemes 5 and 6, where we allocate all power to a single data stream or divide equally to all available data streams. However, when the user is very close to the AP, i.e., \({d_h} < 23\)m, the SNR becomes high; hence, for \({d_h} < 23\)m, it is evident in Figs. \ref{fig:7 SE vs distance Rayleigh} and \ref{fig8:SE vs distance Rayleigh} that Algorithm \ref{alg:highSNR} has improved performance. When the user is located near the DP-IRS, i.e., \({d_h} > 36\)m, we observe similar performance behaviors, as illustrated in Figs. \ref{fig:2SE vs N LoS} and \ref{fig:3SE vs N Rayeligh}.

Finally, in Figs. \ref{fig:9 SE vs antenna vs SPIRS LoS} and \ref{fig:10 SE vs antenna vs SPIRS Rayleigh} for \(\chi = 20\,\rm{dB}\) and \(-20\,\rm{dB}\), respectively, we consider the different numbers of transmit/receive antennas and plot the SE performance of S-IRS and DP-IRS versus the number of reflecting elements. In particular, we consider \({\rm{4}} \times {\rm{4}}\), \({\rm{8}} \times {\rm{8}}\), and \({\rm{16}} \times {\rm{16}}\) MIMO configurations.

\begin{figure}[H]
\centering
\includegraphics[width=0.6\textwidth]{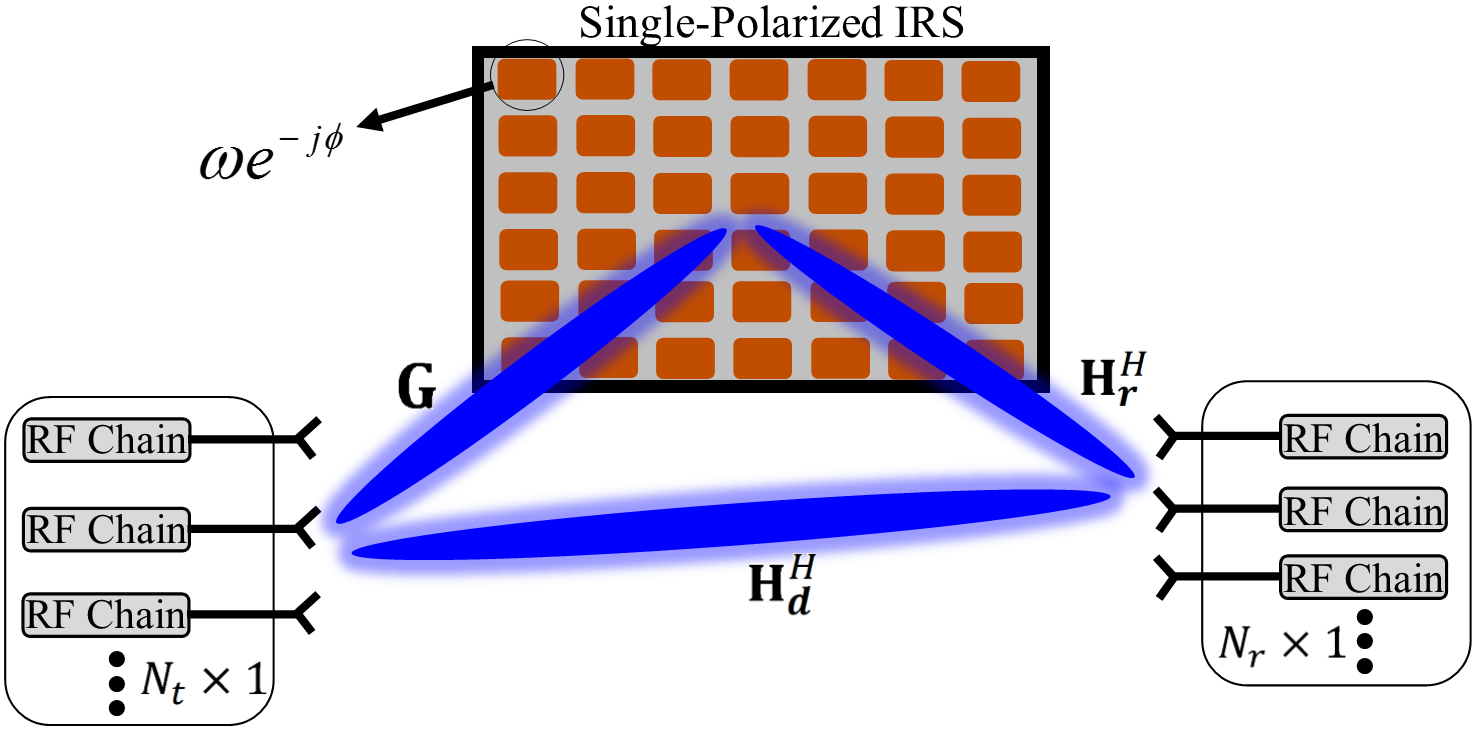}
\caption{\textcolor{black}{System model for the S-IRS-assisted MIMO system.}}
\label{fig:SystemModelSIRS}
\end{figure}

\begin{figure}[H]
\centering
\includegraphics[width=0.5\textwidth]{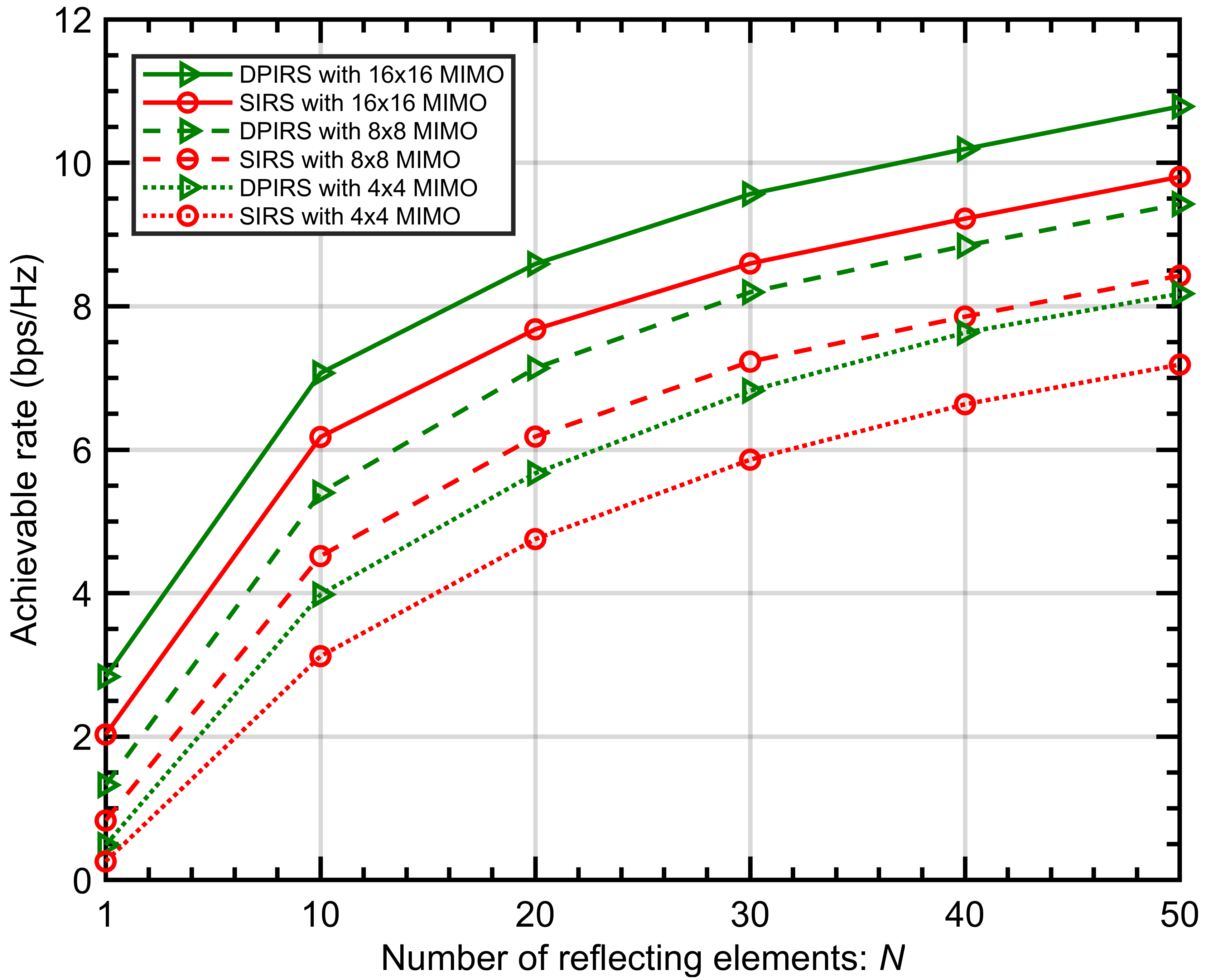}
\caption{Achievable rate of DP-IRS and S-IRS versus the number of reflecting elements for various MIMO configurations (\(\chi = 20\,\rm{dB}\)).}
\label{fig:9 SE vs antenna vs SPIRS LoS}
\end{figure}

\begin{figure}[H]
\centering
\includegraphics[width=0.5\textwidth]{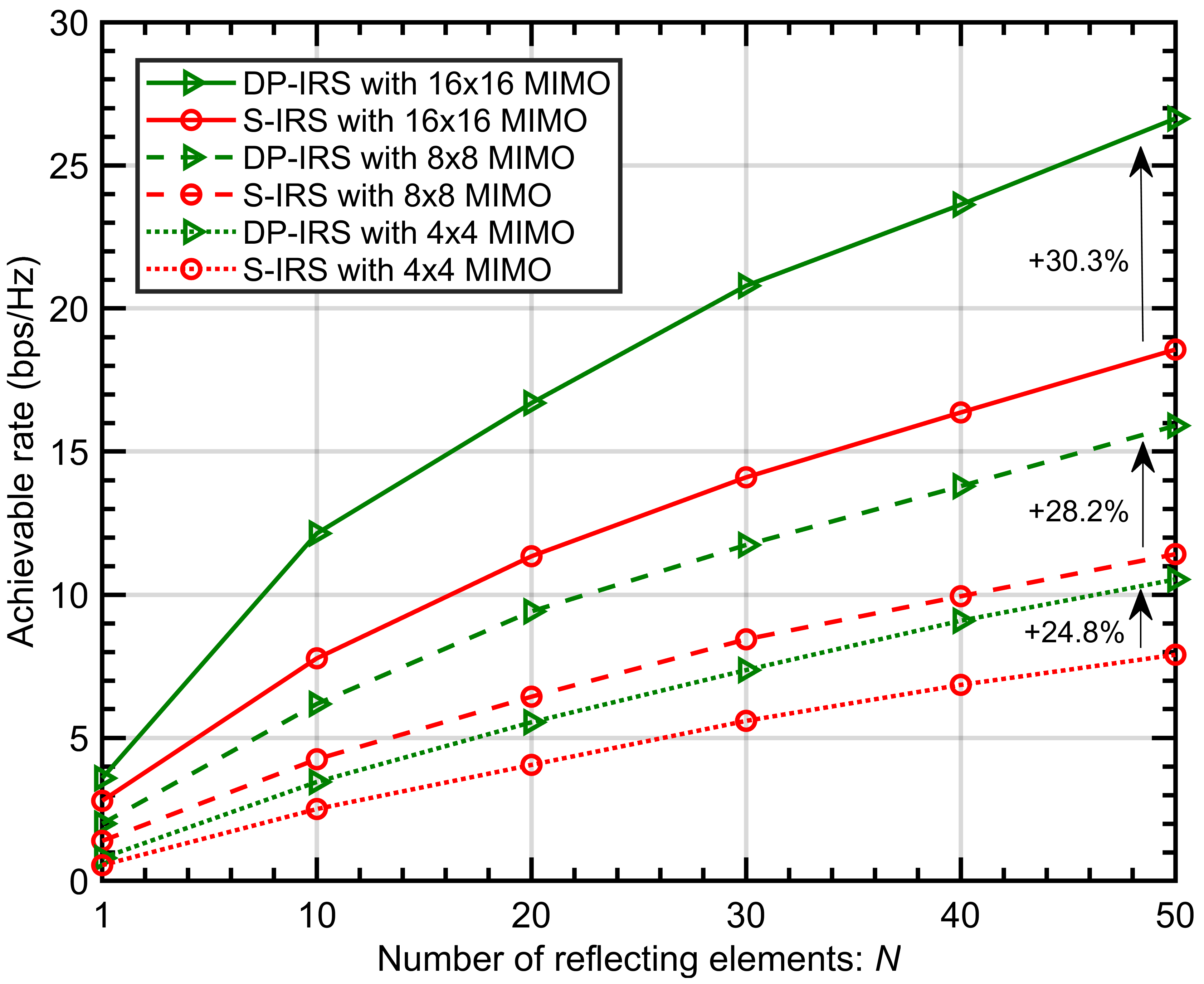}
\caption{Achievable rate of DP-IRS and S-IRS versus the number of reflecting elements for various MIMO configurations (\(\chi = - 20\,\rm{dB}\)).}
\label{fig:10 SE vs antenna vs SPIRS Rayleigh}
\end{figure}

\textcolor{black}{ The system model for the S-IRS-assisted MIMO network is shown in Fig. \ref{fig:SystemModelSIRS}, where each reflecting element of S-IRS introduces phase and amplitude changes in the reflected signals. The operation of a single reflecting element is represented by \(\omega {e^{ - j\phi }}\), where \(\omega \) and \(\phi \) denote the induced amplitude and phase changes, respectively. Similar to our proposed work, we assume full-reflection for each element of the S-IRS, which is achieved by setting \({\omega} = 1\) for all reflecting elements.} For this comparison, we derive a formulation equivalent to \eqref{eq19} for S-IRS phases and devise an AO algorithm that optimizes the IRS phases and precoder/combiner alternatively until convergence is reached. \textcolor{black}{Furthermore, for a fair comparison, we normalize the operation \eqref{eq1} of each reflecting element of the DP-IRS by a factor of \(\frac{1}{{\sqrt 2 }}\), that is, \(\frac{1}{{\sqrt 2 }}{\bf{\Psi }}\).}

Overall, in Figs. \ref{fig:9 SE vs antenna vs SPIRS LoS} and \ref{fig:10 SE vs antenna vs SPIRS Rayleigh}, it can be observed that increasing \(N\), \(N_t\), and \(N_r\) improves the SE performance for both S-IRS and DP-IRS. However, because of the PM gains, DP-IRS performs substantially better than S-IRS. Specifically, in Fig. \ref{fig:10 SE vs antenna vs SPIRS Rayleigh}, for \(\chi = -20\,\rm{dB}\), i.e., Rayleigh fading, more reflecting elements or antennas in the rich scattered environment help DP-IRS to better utilize its PM and PD capabilities, which result in a significantly better SE performance than that of S-IRS. In Fig. \ref{fig:9 SE vs antenna vs SPIRS LoS}, where  \(\chi = 20\,\rm{dB}\), i.e., LoS channel, increasing  \(N\) does not help much in the increase of performance difference between S-IRS and DP-IRS. This is because as long as the dominant channel is LoS, increasing \(N\) does not help much to increase the data streams.

\section{Conclusion}
\label{Sect:Conclusion}

To maximize the SE performance of the DP-IRS-assisted MIMO network, we optimize the operations of the reflecting elements at the DP-IRS, precoder/combiner at the transmitter/receiver, and vertical/horizontal phase shifters at the DP antennas. To solve the formulated capacity maximization problem, we propose an SDR-based AO algorithm, which operates by addressing several subproblems alternately until the SE performance converges. The convex formulations of the sub-problems are shaped, such that the algorithm is guaranteed to converge. Furthermore, considering the low/high SNR regimes, we provide two AO algorithms that are computationally efficient than the first approach. Specifically, for the low-SNR regime, we derive closed-form solutions. We provide extensive numerical results for the SE performance of the DP-IRS by varying the number of reflecting elements, total transmit power, user locations, and type of channels. The numerical results show that the proposed algorithms outperform various benchmark schemes. Specifically, Algorithm 1 achieves a 65.6 \% increase in SE performance compared with random operations and, on average, needs five iterations for convergence. Moreover, a single iteration of Algorithm 1 achieves 83.3 \% of the final converged performance. For the low-SNR regime and/or LoS dominant channel, Algorithm 2 achieves almost the same performance as Algorithm 1 but with a substantially lower computational cost. Specifically, in terms of average CPU running time, for \(N = 50\) and \(N_t = N_r = 16\), Algorithms 1 and 2 take 19.57 s and 10.02 ms, respectively. We also provide the SE
performances of S-IRS and compare them with those of DP-IRS. The numerical results show that for \(N = 50\) and Rayleigh fading, DP-IRS achieves 24.8 \%, 28.2 \%, and 30.3 \% improvements in SE for \({4} \times {4}\), \({8} \times {8}\), and \({16} \times {16}\) MIMO, respectively, compared to the S-IRS. It is noted that the operation of a single reflecting element of the DP-IRS is computationally more expensive than that of a single element of the S-IRS. However, because of the inherent capabilities of DP waves in DP-IRS, we believe that the DP-IRS has much potential over the S-IRS. Correspondingly, an interesting future work could be to investigate the design and performance of the extension of this work to multi-users cases with more practical considerations such as hardware-constrained reflection at DP-IRS, polarization power leakage, imperfect/partial channel state information, etc.


\bibliographystyle{IEEEtran}
\bibliography{IEEEabrv,Main}
\end{document}